\documentclass[12pt,a4paper]{article}
\usepackage[english]{babel}
\usepackage[version=4]{mhchem}
\usepackage{amsmath}
\usepackage{caption} 
\usepackage{nicefrac} 
\usepackage{fontspec}
\usepackage{adjustbox}
\usepackage{subcaption}
\usepackage{graphicx}
\usepackage{lmodern}
\usepackage{exscale}
\usepackage{tikz}
\usepackage{xcolor}
\definecolor{codebg}{RGB}{245,245,245}
\usepackage{xurl}
\usepackage[colorlinks=true, allcolors=blue]{hyperref}
\hypersetup{breaklinks=true}
\usepackage{dblfloatfix}
\usepackage[numbers,sort&compress]{natbib}
\usepackage{amssymb}
\usepackage{physics}
\usepackage{hyperref}
\usepackage{geometry}
\hypersetup{breaklinks=true}
\usepackage{breqn}

\usepackage[title]{appendix} 

\usepackage{listings}
\usepackage{xcolor}

\definecolor{codebg}{RGB}{245,245,245}
\definecolor{codekeyword}{RGB}{0,90,160}      
\definecolor{codestring}{RGB}{170,60,0}       
\definecolor{codecomment}{RGB}{110,110,110}   
\definecolor{codenumber}{RGB}{100,100,180}    
\definecolor{codefunc}{RGB}{100,30,140}       
\definecolor{coderule}{RGB}{200,200,200}      

\lstdefinestyle{python}{
  language=Python,
  backgroundcolor=\color{codebg},
  basicstyle=\ttfamily\small,
  keywordstyle=\color{codekeyword}\bfseries,
  commentstyle=\color{codecomment}\itshape,
  stringstyle=\color{codestring},
  numberstyle=\tiny\color{codecomment},
  identifierstyle=\color{black},
  emph={self,__init__,None,True,False},
  emphstyle=\color{codefunc}\bfseries,
  breaklines=true,
  breakatwhitespace=false,
  showstringspaces=false,
  columns=fullflexible,
  keepspaces=true,
  frame=single,
  rulecolor=\color{coderule},
  framerule=0.4pt,
  framesep=6pt,
  xleftmargin=8pt,
  xrightmargin=8pt,
  tabsize=4,
  captionpos=b,
  numbers=none,           
  numbersep=8pt,
  belowskip=8pt,
  aboveskip=8pt,
}

\lstdefinestyle{bash}{
  language=bash,
  backgroundcolor=\color{codebg},
  basicstyle=\ttfamily\small,
  keywordstyle=\color{codekeyword}\bfseries,
  commentstyle=\color{codecomment}\itshape,
  stringstyle=\color{codestring},
  breaklines=true,
  breakatwhitespace=false,
  showstringspaces=false,
  columns=fullflexible,
  keepspaces=true,
  frame=single,
  rulecolor=\color{coderule},
  framerule=0.4pt,
  framesep=6pt,
  xleftmargin=8pt,
  xrightmargin=8pt,
  tabsize=4,
  numbers=none,
  belowskip=8pt,
  aboveskip=8pt,
}

\usetikzlibrary{shadows,positioning,calc}

\title{ASE2SPRKKR: a unified Python framework integrating the Spin-Polarized Relativistic Korringa-Kohn-Rostoker method into the Atomic Simulation Environment}
\date{}

\author{
Ridha Eddhib$^{1}$, Matyáš Novák$^{1,2}$, Hubert Ebert$^{3}$,\\ Aki Pulkkinen$^{1}$, Ján Minár$^{1,*}$}

\date{}

\begin{document}
\maketitle
\renewcommand{\thefootnote}{\fnsymbol{footnote}}
\begin{center}
{\small
$^{1}$ New Technologies-Research Centre, University of West Bohemia in Pilsen, 30100 Pilsen, Czech Republic \\
$^{2}$ FZU – Institute of Physics, Czech Academy of Sciences, v.v.i, Prague, Czech Republic \\
$^{3}$ Department Chemie, Ludwig-Maximilians-Universität München, Butenandtstr. 5-11, 81377 München, Germany
}
\footnote{jminar@ntc.zcu.cz} 
\end{center}

\begin{abstract}
The Spin-Polarized Relativistic Korringa-Kohn-Rostoker (SPR-KKR) is an all-electron ab-initio multiple-scattering code that provides unique capabilities for treating chemical disorder, finite-temperature magnetism, relativistic effects, and spectroscopic properties of various types of solids through its fundamental formulation in terms of the single-particle Green's function rather than eigenstates. We present ASE2SPRKKR, a comprehensive Python interface that integrates SPR-KKR into the Atomic Simulation Environment, making SPR-KKR more accessible, streamlined, and uniform. 
Our implementation extends the Atomic Simulation Environment (ASE)'s \texttt{Atoms} object to handle fractional site occupations for coherent-potential-approximation calculations while maintaining full compatibility with ASE's extensive ecosystem of structure builders, optimizers, and analysis tools. Automated input generation with validation, comprehensive output parsing, and direct MPI support enable seamless integration into high-throughput and multi-method workflows. We demonstrate the interface through representative applications: semi-infinite surface calculations reproducing Rashba-split Au(111) surface states; one-step photoemission modeling capturing matrix-element effects; exchange-parameter extraction for atomistic spin dynamics; and X-ray absorption spectroscopy including magnetic circular dichroism. Beyond these demonstrations, ASE2SPRKKR is designed with transferability as a first-class concern. By grounding its architecture in FAIR principles of Findability, Accessibility, Interoperability, and Reusability, it establishes a replicable blueprint for bringing other specialized Green's function and first-principles codes into the collaborative, reproducible workflows that modern materials discovery requires.

\end{abstract}


\section{Introduction}

Computational materials science relies predominantly on density functional theory (DFT) \citep{Hohenberg1964, Kohn1965}, which maps the many-body electronic problem onto a non-interacting Kohn-Sham formulation. As computational power continues to grow, a key challenge for computational physics is to exploit high-performance computing resources while providing flexible interfaces for setting up and managing sophisticated calculations across multiple codes.

The Atomic Simulation Environment (ASE) \citep{Larsen2017, Bahn2002} has emerged as the leading Python framework for atomistic simulations, currently interfacing with over 30 electronic structure codes. Most ASE-supported codes employ eigenvalue-based methods, plane-wave \citep{Kresse1996, Giannozzi2009, Gonze2020}, localized-orbital \citep{Soler2002, Kuhne2020, Blum2009}, or augmented-wave approaches \citep{Blaha2020, Wortmann2023, Enkovaara2010}, solving for single-particle eigenfunctions $\psi_n(\mathbf{r})$ and eigenvalues $\varepsilon_n$ to construct electron densities and total energies. While remarkably successful for ordered crystalline systems, these variational methods face fundamental challenges when treating broken translational symmetry, substitutional disorder \citep{Soven1967, Gyorffy1972, Zunger1990}, or computing response functions and spectroscopic observables \citep{Runge1984, Onida2002, Ebert2011}.

An alternative formulation based on the single-particle Green's function $G(\mathbf{r},\mathbf{r}';E)$ \citep{Zeller1995, Faulkner1980} addresses these limitations. The Korringa-Kohn-Rostoker (KKR) method \citep{Korringa1947, Kohn1954}, formulated through multiple scattering theory \citep{Gonis2000, Ebert2011}, constructs the Green's function as the central computed quantity rather than deriving it from eigenstates. While mathematically equivalent to eigenvalue methods for perfect crystals, this reformulation provides decisive computational and conceptual advantages for several problem classes that remain challenging for conventional approaches \citep{Turek1997}.

For substitutional disorder in alloys $A_xB_{1-x}$, the coherent potential approximation (CPA) \citep{Soven1967, Taylor1967, Gyorffy1972} replaces configurational averaging \citep{Turek1997, Faulkner1980} over random atomic
distributions via an effective medium determined self-consistently. Supercell and special quasirandom structure (SQS) approaches \citep{Zunger1990, Magri1991} face fundamental scaling limitations: representing a minority species at concentration $x$ requires $\mathcal{O}(x^{-1})$ atoms, and since DFT cost scales cubically with system size, the total cost grows as $\mathcal{O}(x^{-3})$ \citep{Ruban2008, Abrikosov2016} in the unit cell, a $10^6$-fold penalty at $x=0.01$ relative to the equiatomic case. SQS cells carry an additional burden of reproducing multi-body correlation functions, becoming intractable for multicomponent or dilute systems \citep{Alling2010}. The single-site CPA \citep{Ebert2011} circumvents these issues entirely, with computational cost strictly $\mathcal{O}(1)$ in $x$. 
Furthermore, extensions to non-local CPA \citep{Rowlands2003, Jarrell2001} systematically recover short-range order effects beyond the single-site approximation.

The disordered local moment (DLM) approach \citep{Gyorffy1985, Staunton2014} maps finite-temperature magnetic fluctuations onto orientational disorder, which is treatable within the CPA framework, enabling \textit{ab initio} predictions of paramagnetic properties and Curie temperatures. Combined with temperature-dependent lattice vibrations through alloy-analogy models, this provides comprehensive access to realistic finite-temperature electronic structure.

In spectroscopy, the Green's function formalism \citep{Economou1983, Fetter1971, Ebert2011} provides a natural connection to experimental observables. Photoemission spectroscopy directly measures quantities proportional to the spectral function \citep{Hufner2003, Damascelli2003, Lv2019}. The one-step photoemission model \citep{Pendry1976, Braun1996, Braun2006, Minar2020} treats excitation, photoelectron propagation, and detection as a single coherent quantum process, capturing matrix-element effects \citep{Fadley1992, Schattke2003} and photoelectron diffraction essential for the quantitative interpretation of angle-resolved photoemission (ARPES) experiments \citep{Reinert2005, Lv2019}. X-ray absorption and magnetic circular dichroism calculations \citep{Stohr1999, Ankudinov1998, Ebert1996, Arola1997} benefit from the all-electron treatment of core states combined with a fully relativistic formalism \citep{Rose1961, Strange1998} incorporating spin-orbit coupling at all orders.

Transport properties emerge naturally through the Kubo-Green's function formalism \citep{Kubo1957, Bastin1971, Crepieux2001}, expressing conductivity tensors in terms of Green's function and the corresponding matrix elements without requiring explicit eigenstate construction. This enables efficient calculations of electrical conductivity \citep{Lowitzer2010, Butler1985}, thermoelectric coefficients \citep{Streda1982}, anomalous Hall effect \citep{Lowitzer2011, Nagaosa2010}, and quantum transport \citep{Landauer1957, Buttiker1986, Datta1995} in disordered systems where Bloch-state methods struggle.

Multi-scale magnetic modeling benefits from direct parameter extraction capabilities. The Lichtenstein formula \citep{Liechtenstein1987, Udvardi2003, Ebert2009} expresses the Heisenberg exchange parameters $J_{ij}$ as energy integrals of Green's functions, avoiding computationally expensive total-energy differences. Gilbert damping calculations \citep{Gilbert2004, Brataas2008, Ebert2011b, Mankovsky2013} based on torque-correlation functions yield dissipation parameters for magnetization dynamics. These \textit{ab initio} parameters feed directly into atomistic spin dynamics and Monte Carlo simulations \citep{Skubic2008, Evans2014, Eriksson2017, Bergman2010}, bridging electronic structure to mesoscale magnetic phenomena.

The Green's function serves as the central object in diagrammatic many-body perturbation theory, facilitating systematic extensions beyond standard DFT. The fully self-consistent KKR+DMFT framework \citep{Georges1996, Kotliar2006, Minar2005, Minar2011} combines the KKR-GF method with dynamical mean-field theory (DMFT), introducing frequency-dependent self-energies obtained by local quantum impurity solvers. The charge self-consistency between the KKR host and the DMFT solver ensures proper feedback between local correlations and itinerant band states, enabling \textit{ab initio} predictions of spectroscopic properties in strongly correlated systems such as transition-metal oxides and heavy-fermion compounds \citep{Imada1998, Held2007}, where standard DFT fails.

The spin-polarized relativistic KKR (SPR-KKR) code \citep{Ebert2000, EbertSPRKKR} implements these methodologies within a mature, extensively validated framework. Comprehensive reviews \citep{Ebert2011, Gonis2000, zabloudil2005} detail its theoretical foundations and applications. Despite proven capabilities and impact, SPR-KKR has remained outside the ASE ecosystem, while other KKR implementations have already been integrated into Python-based workflow frameworks \citep{aiidakkr}. This limits its accessibility for researchers who would benefit from automated workflows, integration with complementary codes, and modern Python-based analysis tools.

\begin{figure}
  \centering
  {\fontsize{4pt}{5pt}
  \includegraphics[width=0.95\textwidth]{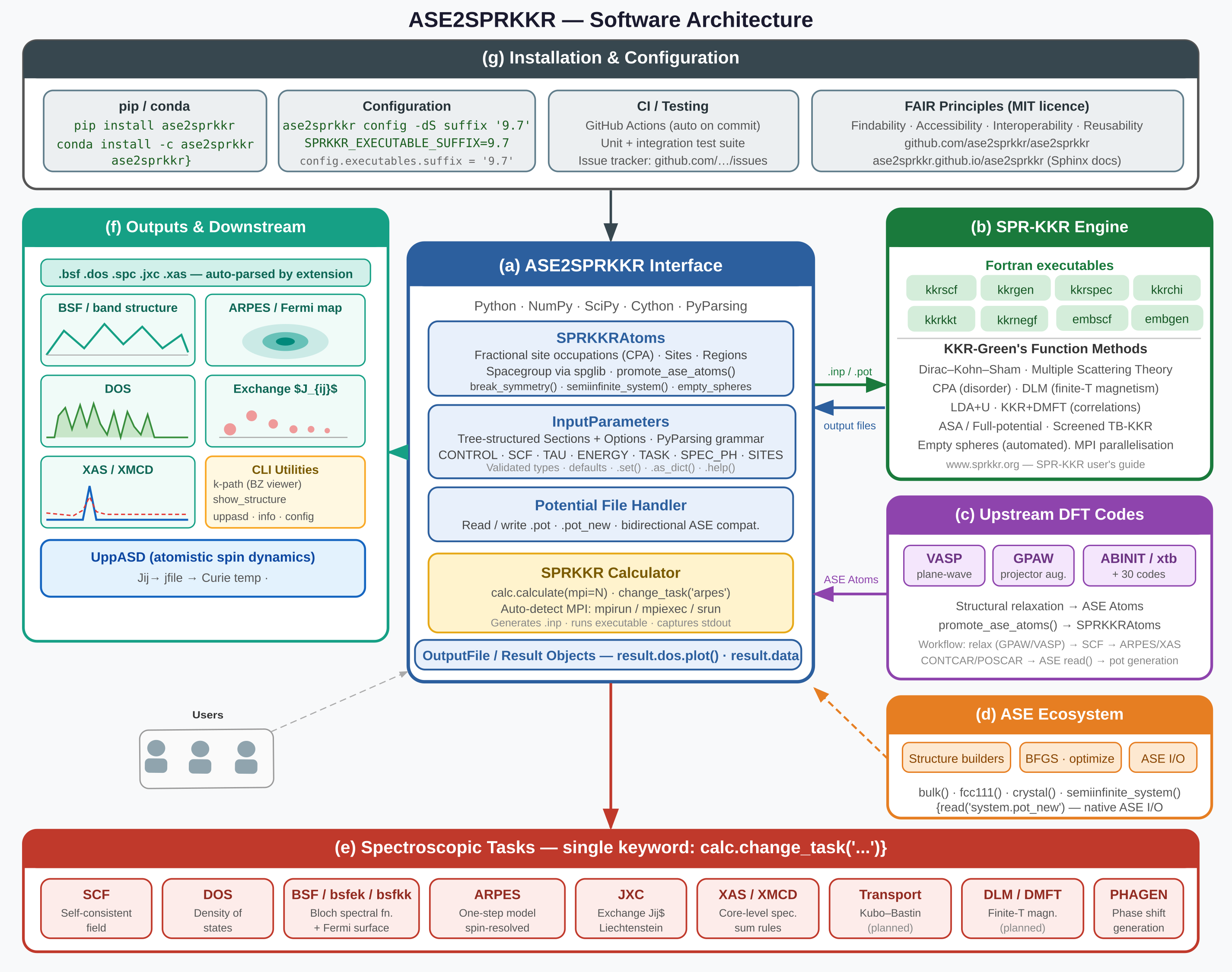}
  \caption{ \textbf{Software architecture of ASE2SPRKKR.} \textbf{(a)}~Core interface: \texttt{SPRKKRAtoms} extends ASE  \texttt{Atoms} with fractional site occupations (CPA), spglib symmetry, and \texttt{Region} support; \texttt{InputParameters} provides a validated, tree-structured parameter hierarchy (PyParsing grammar); the \texttt{Potential File Handler} manages bidirectional \texttt{.pot}/\texttt{.pot\_new} I/O; the \texttt{SPRKKR Calculator} drives input generation, MPI execution, and task switching via \texttt{change\_task()}; structured \texttt{Result} objects expose arrays and built-in \texttt{plot()} methods.
    \textbf{(b)}~SPR-KKR Fortran backend~\cite{Ebert2011}(\texttt{kkrscf}/\texttt{kkrgen}/\texttt{kkrspec}/\texttt{kkrchi}/\texttt{kkrkkt}/\texttt{kkrnegf}/\texttt{embscf}/\texttt{embgen}): CPA~\cite{Soven1967}, DLM~\cite{Gyorffy1985}, LDA$+U$~\cite{Liechtenstein1995}, KKR$+$DMFT~\cite{Minar2011}, ASA/full-potential, screened TB-KKR~\cite{Szunyogh1994}, and layer doubling~\cite{MacLaren1990}.
    \textbf{(c)}~Upstream DFT codes (VASP, GPAW, ABINIT, 30+): relaxed \texttt{Atoms} structures are promoted to \texttt{SPRKKRAtoms} via \texttt{promote\_ase\_atoms()}
    without manual file conversion.
    \textbf{(d)}~ASE ecosystem~\cite{Larsen2017}: structure builders, BFGS optimiser, and native \texttt{.pot\_new} I/O. 
    \textbf{(e)}~Available tasks via \texttt{calc.change\_task()}:SCF, DOS, BSF/Fermi surface, ARPES (one-step model~\cite{Pendry1976,Braun1996}), exchange $J_{ij}$ (JXC~\cite{Liechtenstein1987}), XAS/XMCD with sum rules~\cite{Thole1992,Carra1993}, DLM/DMFT and PHAGEN; transport workflows (Kubo-Bastin~\cite{Butler1985}) are planned.
    \textbf{(f)}~Outputs and downstream: parsed result objects, CLI utilities (\texttt{k-path}, \texttt{plot}), and one-command $J_{ij}$ export to UppASD~\cite{Eriksson2017} for Curie temperatures.
    \textbf{(g)}~Distribution (pip/conda), executable configuration, GitHub Actions CI, and FAIR principles
    compliance (MIT licence, Sphinx docs, public issue tracker).
  \label{fig:architecture}}}
\end{figure}

We address these limitations through ASE2SPRKKR, which integrates SPR-KKR within ASE's infrastructure while preserving its specialized capabilities. The overall
architecture of ASE2SPRKKR is illustrated in Fig.~\ref{fig:architecture}. ASE2SPRKKR handles KKR-specific requirements: flexible occupation specifications for CPA disorder; cached space-group symmetry for irreducible-wedge calculations; Region objects for tight-binding embedding and layered geometries. Implementation via \texttt{SPRKKRAtoms} objects extends the standard ASE \texttt{Atoms} class without breaking its base functionality, ensuring bidirectional compatibility. Expert users retain direct access to potential and task files, while simplified interfaces support common workflows. Input validation prevents setup errors; output parsing extracts densities of states, spectral functions, transport coefficients, and magnetic properties into standard Python structures.

This integration enables hybrid workflows combining complementary methods: structural relaxations with other codes followed by spectroscopic calculations with SPR-KKR's one-step photoemission model; high-throughput screening of ordered structures followed by systematic disorder and temperature exploration via CPA and DLM; exchange parameter extraction followed by spin dynamics parameterization. ASE2SPRKKR substantially lowers barriers for exploiting the unique capabilities developed over decades, particularly for problems involving disorder, magnetism, relativistic effects, and spectroscopy, where the KKR-GF framework excels. The ongoing developments aim to further streamline workflows, incorporate additional SPR-KKR features, and enhance the accessibility and versatility of this powerful framework for the broader scientific community. 

This paper is organized as follows. Section~\ref{sec:software} outlines the software architecture, core components, and integration patterns. Section~\ref{sec:applications} demonstrates the capabilities of our approach through representative applications and performance analysis. We conclude with a discussion of current implementations and the development road map. All calculations employed SPR-KKR version 9.7. The supplementary material  provides the theoretical foundations and SPR-KKR implementation details necessary for understanding the interface design.

\section{ASE2SPRKKR Software Architecture}
\label{sec:software}

Python~\citep{python} has become one of the most widely used programming languages in the scientific community, primarily due to its extensive ecosystem of libraries for numerical and scientific computation. Among these, \texttt{NumPy}~\citep{harris2020array} provides efficient numerical array manipulation, \texttt{SciPy}~\citep{virtanen2020scipy} implements a wide variety of scientific algorithms including optimization, integration, and interpolation, and \texttt{Cython}~\citep{behnel2010cython} enables high-performance interoperability between Python and compiled languages such as C or Fortran. Given this rich ecosystem and its flexibility, Python was a natural choice for developing the ASE2SPRKKR interface.

To facilitate interoperability with other atomistic simulation codes and minimize implementation effort, ASE2SPRKKR is built on the Atomic Simulation Environment (ASE)~\citep{Larsen2017}. ASE provides a standardized and extensible framework for describing, manipulating, and visualizing atomic structures directly within Python. It offers a unified interface for setting up systems, running simulations, and analyzing results using a wide range of back-end calculators—from \texttt{ABINIT} to \texttt{VASP} and \texttt{xtb}. ASE2SPRKKR serves as a bridge that allows users to perform spin-polarized and relativistic electronic structure calculations using the SPR-KKR package directly within ASE. The complete ASE documentation is available online at \url{https://wiki.fysik.dtu.dk/ase/}.

ASE2SPRKKR does \emph{not} reimplement SPR-KKR functionality. It generates input files, manages execution, and parses outputs, maintaining complete compatibility with native SPR-KKR. Users can mix ASE2SPRKKR workflows with manual input file editing, useful for experimental features or troubleshooting.

\subsection{Installation and Configuration}

Installation of ASE2SPRKKR follows standard Python package management practices. The recommended approach uses pip (Package Installer for Python), which is the standard way to install Python packages:

\begin{lstlisting}[style=bash]
pip install ase2sprkkr
\end{lstlisting}

Using a Python virtual environment is highly recommended but not required. Alternatively, the Anaconda distribution system~\citep{anaconda} can be used:

\begin{lstlisting}[style=bash]
conda install -c ase2sprkkr ase2sprkkr
\end{lstlisting}

ASE2SPRKKR is under active development. The stable release version can be supplemented with the beta version, which often includes new features and bug fixes:

\begin{lstlisting}[style=bash]
pip install --pre ase2sprkkr
\end{lstlisting}

For cutting-edge features, the development branch of the Git repository can be accessed directly:

\begin{lstlisting}[style=bash]
 pip install git+https://github.com/ase2sprkkr/ase2sprkkr.git@develop
\end{lstlisting}

To run SPRKKR calculations with ASE2SPRKKR, the SPR-KKR executables must be compiled and available in the ASE2SPRKKR configuration file. The configuration file path can be displayed using:

\begin{lstlisting}[style=bash]
ase2sprkkr config -p
\end{lstlisting}

The executable suffix is set directly via:

\begin{lstlisting}[style=bash]
ase2sprkkr config -dS executables.suffix '9.7'
\end{lstlisting}

For workflows requiring multiple SPR-KKR versions, the suffix can be modified programmatically in calculation scripts:

\begin{lstlisting}[style=python]
from ase2sprkkr.configuration import config 
config.executables.suffix = '9.7'
\end{lstlisting}

or through environment variables:

\begin{lstlisting}[style=bash]
export SPRKKR_EXECUTABLE_SUFFIX=9.7
\end{lstlisting}

SPR-KKR executables should be in the \texttt{PATH} environment variable. If not, the executable directory can be specified in the \texttt{executables.dir} configuration variable. Additional configuration options are accessible via:

\begin{lstlisting}[style=bash]
ase2sprkkr config -h
\end{lstlisting}

\subsection{Extended Atomic Structure: SPRKKRAtoms}
\label{subsec:sprkkratoms}

ASE~\citep{Larsen2017} is a powerful and versatile software package that supports a wide range of codes, from ab-initio calculations to molecular dynamics. However, support for certain properties specifically required by SPR-KKR is either limited, not very user-friendly, or entirely missing. For example, occupation is represented only as an array of occupation dictionaries associated with specific site-types. Consequently, if one wants to change or break the symmetry of a structure, the entire array has to be rebuilt. While this is certainly possible, it sacrifices the elegance and simplicity of a scripting interface that is accessible even to users with limited programming experience. To address this and allow typical users to easily run SPR-KKR calculations, the functionality of ASE \texttt{Atoms} objects has been extended through \texttt{SPRKKRAtoms}.

To allow a smooth transition from ASE to ASE2SPRKKR and back, the method \texttt{promote\_ase\_atoms} has been introduced. This method adds an additional parent class (in the object-oriented programming sense) to the existing ASE Atoms object.
The promotion does not alter the original object in any other way: it can still be used in the ASE-way as before, it just provides additional functionality. Extension also occurs automatically when \texttt{Atoms} interfaces with ASE2SPRKKR functions, maintaining full backward compatibility:
\label{Ni2FeGa.example}
\begin{lstlisting}[style=python]

from ase2sprkkr import SPRKKRAtoms
from ase.visualize import view 
import numpy as np 

# Create ordered Heusler structure
# 1. Define primitive FCC cell vectors
primitive_cell = np.array([
    [0.0, 0.5, 0.5],
    [0.5, 0.0, 0.5],
    [0.5, 0.5, 0.0]
]) * 5.748

# 2. Define the scaled positions
scaled_positions = [
    [0.25, 0.25, 0.25], # Ni (1)
    [0.75, 0.75, 0.75], # Ni (2)
    [0.5,  0.5,  0.5],   # Fe
    [0.0,  0.0,  0.0],   # Ga
]

# 3. Build the Atoms object directly
Ni2FeGa = SPRKKRAtoms(
    symbols=['Ni', 'Ni', 'Fe', 'Ga'],
    scaled_positions=scaled_positions,
    cell=primitive_cell,
    pbc=True
)

# 4. Define Fe/Co disorder
Ni2FeGa.sites[2].site_type.occupation.set({'Fe':0.75, 'Co':0.25})
# 5. View and save the structure
view(Ni2FeGa,viewer='avogadro',repeat=[2,2,2])
Ni2FeGa.potential.save_to_file("Ni2FeGa.pot")

\end{lstlisting}

\paragraph{Sites}
The \texttt{sites} property, which is in fact stored as an ASE array to retain smooth manipulation with \texttt{SPRKKRAtoms} objects via summing and slicing, provides a user-friendly way to specify atomic sites, their occupations, and to associate additional SPR-KKR-specific properties, such as radial potentials or charges.

 Site holds information about magnetization of the given site and groups symmetry-equivalent sites using reference to a shared \texttt{SiteType} object.
 
\texttt{SiteType} object carries:
\begin{itemize}
    \item \textbf{Occupation dictionary}: maps species to fractional occupations, summing to unity
    \item \textbf{Radial potential and charge}: result of a self-consistent calculation (once performed)
    \item \textbf{Radial mesh}: for the potential and charge
\end{itemize}
Moreover \texttt{Site} and \texttt{SiteType} allows user-friendly manipulation with the sites, as breaking symmetry, 
changing occupation etc.

Some properties of a \texttt{site} or \texttt{site\_type} duplicate the original ASE properties (e.g., \texttt{symbols} array, occupation array). When a \texttt{site} property is edited, the change is mirrored to the corresponding ASE property to maintain backward compatibility. However, the reverse is not true due to the inherent limitations of ASE. Therefore, modifications of ASE properties, such as the \texttt{symbols} of \texttt{Atoms} objects, are recommended only before promotion.

When saving the potential, e.g. at the beginning of the \texttt{calculate} or \texttt{save\_input} methods of \texttt{calculator}, the values of \texttt{SPRKKRAtoms} are written to the \texttt{potential} using the \texttt{set\_from\_atoms} methods of sections. In contrast, reading potentials, or manually updating atoms via the \texttt{update\_atoms} method of a \texttt{Potential}, is done using the \texttt{update\_atoms} methods.\footnote{In the current state of development, not all sections of \texttt{potential} are parsed; some are just stored as raw data and retained in the \texttt{potential} object.} In this way, the original structure of \texttt{potential} remains accessible, allowing low-level manipulations with \texttt{potential} data if needed, while enabling user-friendly editing of the common properties of the calculated structure via the \texttt{SPRKKRAtoms} object and its \texttt{sites} property.

\paragraph{Regions}
SPR-KKR supports tight-binding mode calculations, in which two 3D bulk materials are connected by a transition region that is periodic in two dimensions. For this purpose, the ASE \texttt{Atoms} object is extended with \texttt{Region} objects (accessible via \texttt{[add_|set_|remove_]regions} methods).

Each \texttt{Region} has its own cell, periodicity, and symmetry information, as well as a mask specifying which sites belong to the region.

\paragraph{Spacegroup information}
SPR-KKR requires information about the symmetry of atoms and regions. Spglib~\citep{spglib} automatically determines space group symmetry (space group 225, F$\overline{4}$3m for the Heusler structure above), linking symmetry-equivalent sites. Since computing this information can be computationally expensive, the results are stored in the \texttt{spacegroup\_info} property of the \texttt{SPRKKRAtoms} object. This information is used during initialization of the \texttt{sites} array and when building the potential just before a calculation.

\subsection{ASE2SPRKKR Data Objects}
\label{subsec:dataobjects}

Many of the ASE2SPRKKR objects, for example \texttt{InputParameters}, \texttt{Potential}, and all \texttt{OutputFile}, share the same implementation: a tree-like data.
Each item of the data-tree has an associated \emph{definition class}, an approach similar to the way in which, e.g., the well-known NOMAD~\citep{Scheidgen2023} organizes its data on the Python level.\footnote{However, the implementation differs, since NOMAD uses metaclasses~\citep{Scheidgen2023} for this purpose (which has its pros and cons), while ASE2SPRKKR describes the structure of the data using regular Python classes.}

Definition classes describe both the structure of data and the way in which the data is stored in a file. Each section and option value of an \texttt{InputFile} has a property \texttt{definition}. The definition of a section states which are the members of the section, and the rules for how to read/write them: i.e., how they are separated (whether by newlines, spaces, tabs, etc.), whether they have a fixed order or not, and so on. The definition of a value specifies the data type of the value, whether it is required, and the default value for the option. The definitions also contain documentation for the data.

Data types are described by descendants of the \texttt{GrammarTypes} object; these objects describe the values and their format in the resulting file. They can be nested: one option can hold more than one value, either resulting in a NumPy array~\citep{harris2020array} or in a tuple of different types, or even 2D tables in the data (see the example in Fig.~\ref{Fig:lattice}).

Saving the data into a file is straightforward: the data tree is traversed, and each definition takes care of saving itself and its child objects. Parsing is more involved: for this purpose, the definition generates a grammar using PyParsing~\citep{mcguire2007getting}. Using the grammar, the file is parsed, resulting in a dictionary that is used to set up the resulting object.

Using this mechanism, defining new sections within a \texttt{Potential}, adding new values to \texttt{InputParameters}, or introducing a new task type becomes very easy. Figure~\ref{Fig:lattice:def} shows an example of how a Potential section can be defined to describe the unit cell of a material being calculated. As demonstrated in Fig.~\ref{Fig:lattice:vis}, defining even a relatively complex nested structure is straightforward. Example data that such a section may contain are shown in Fig.~\ref{Fig:lattice:data} in their raw text form, and Fig.~\ref{Fig:lattice:python} presents their Python representation.

\subsection{Calculator Interface and Execution Model}
Following ASE’s calculator paradigm, computation proceeds in three steps: instantiation, parameter configuration, and execution (performed via the \texttt{calculate} method).

\begin{lstlisting}[style=python]
from ase2sprkkr import SPRKKR

calculator = SPRKKR(atoms=Ni2FeGa)
calculator.input_parameters.SCF.NITER = 300
calculator.input_parameters.TAU.NKTAB = 150
calculator.calculate()
\end{lstlisting}

The \texttt{calculate()} method:
\begin{enumerate}
\item Generates potential file from \texttt{atoms} structure
\item Writes task-specific input file from \texttt{input\_parameters}
\item Invokes SPR-KKR executable with MPI if available
\item Monitors execution, capturing output
\item Parses results into task-dependent result objects
\item Returns result object with computed properties
\end{enumerate}

MPI parallelization auto-detection queries common MPI implementations~\citep{mpi1994} (\texttt{mpirun}, \texttt{mpiexec}, or Slurm HPC Workload manager \cite{2003_Yoo, slurm_manual} \texttt{srun}). 

Explicit control supports custom configurations:
\begin{lstlisting}[style=python]
calculator.calculate(mpi=16)  # 16 processes
calculator.calculate(mpi=['/opt/mpi/bin/mpirun','-n','16'])
\end{lstlisting}

\subsection{Input Parameter Management}

The \texttt{InputParameters} object provides hierarchical access to task-specific parameters organized by sections (CONTROL, SCF, TAU, ENERGY, TASK, etc.). Each parameter includes type validation, range constraints, and documentation:

\begin{lstlisting}[style=python]
# Task selection modifies available parameters
calculator.change_task('arpes')

# Batch parameter setting with validation
calculator.input_parameters.set({
    'ENERGY.EMINEV': -4.0,
    'ENERGY.EMAXEV': 2.0,
    'ENERGY.NE': 300,
    'SPEC_PH.EPHOT': 50.0,
    'TASK.MILLER_HKL': [1,1,1],
})

\end{lstlisting}

Parameter values can be extracted for further processing:

\begin{lstlisting}[style=python]
calculator.input_parameters.as_dict()
\end{lstlisting}

Documentation of available parameters is provided in the SPR-KKR manual~\citep{EbertSPRKKR}. Brief parameter descriptions are accessible from the command line:

\begin{lstlisting}[style=bash]
ase2sprkkr info --task                # show available tasks
ase2sprkkr info --task=scf            # show SCF task sections
ase2sprkkr info --task=scf.control    # show CONTROL section parameters
\end{lstlisting}

or in interactive Python:

\begin{lstlisting}[style=python]
from ase2sprkkr import SPRKKR
calculator = SPRKKR(task='scf')
calculator.input_parameters.CONTROL.help()
\end{lstlisting}

Input parameters can be read from files to reproduce or modify previous computations:

\begin{lstlisting}[style=python]
from ase2sprkkr import InputParameters
calculator.input_parameters = \
    InputParameters.from_file('previous.inp')
\end{lstlisting}

Input parameter properties are validated to prevent setting parameters to invalid values. However, ASE2SPRKKR may not yet support certain parameters, or validation may be overly strict. In these cases, restrictions can be overridden:

\begin{lstlisting}[style=python]
# Add unsupported parameter
ip.CONTROL.add('MY_NEW_PARAMETER', its_value)
# Set parameter to unsupported value
ip.SCF.NITER.set_dangerous('infinity')
\end{lstlisting}

These should be used sparingly, as they bypass safety checks.

\subsection{Potential File Handling}
Potential files encode the system structure, including the lattice, atomic positions, site types, and their occupations. 
After the user runs a self-consistent calculation (SCF task), a new \texttt{original\_file.pot\_new} file is generated, which also contains the results of the calculation, such as potentials, charge densities, and magnetic moments. In some advanced workflows, such a “converged” potential can also be created manually.

Non-SCF tasks (DOS, ARPES, BSF, etc.) require  such converged potentials:
\begin{lstlisting}[style=python]
calc = SPRKKR()
result = calc.calculate(
    potential='converged.pot_new',
    input_parameters='dos'
)
\end{lstlisting}

Potentials are compatible with ASE in both directions and can be read using either the ASE2SPRKKR API:
\begin{lstlisting}[style=python]
from ase2sprkkr import Potential
pot = Potential.from_file('system.pot_new')
atoms = pot.atoms
print(atoms.symbols)
\end{lstlisting}

or using ASE's native \texttt{read} method:
\begin{lstlisting}[style=python]
from ase.io import read
atoms = read('system.pot_new')
magnetic_moments = [site.moments for site in atoms.sites]
\end{lstlisting}
This enables workflows where structural optimizations with other codes precede spectroscopic calculations with SPR-KKR.

\subsection{Output Processing and Visualization}

Task-dependent result objects provide property access with built-in visualization:
\begin{lstlisting}[style=python]
# DOS calculation
result = calculator.calculate(input_parameters='dos')
result.dos.plot()                # matplotlib visualization
energies = result.dos.EFERMI()   # Fermi energy
total_dos = result.dos.ENERGY()  # Energy grid as NumPy array
\end{lstlisting}

Independent file parsing supports post-processing:
\begin{lstlisting}[style=python]
from ase2sprkkr import OutputFile
bsf = OutputFile.from_file('calc.bsf')
bsf.plot(sites=[0,1])  # plot specific atomic layers
bsf.data  # access raw data arrays
\end{lstlisting}

File type inference from extensions (.bsf, .dos, .spc, .jxc, .xas) or content analysis automatically selects  the appropriate parser.

\subsection{Workflow Integration Patterns}

ASE2SPRKKR is designed to interoperate with the broader ASE ecosystem and external codes, enabling several practical integration patterns for multi-step computational studies.

\paragraph{Sequential multi-method calculations:}
A common workflow combines a plane-wave DFT code for structural relaxation with SPR-KKR for spectroscopic post-processing. Since ASE2SPRKKR operates on standard ASE \texttt{Atoms}
objects, the relaxed geometry produced by any ASE-compatible calculator can be passed directly to SPR-KKR without manual file conversion. In the example below, GPAW is used to obtain the equilibrium geometry, after which the optimized structure is forwarded to SPR-KKR for an ARPES calculation:
\begin{lstlisting}[style=python]
from ase.spacegroup import crystal
from ase.filters import ExpCellFilter
from ase.optimize import BFGS
from ase.io import write
from gpaw import GPAW, PW, FermiDirac
from ase.filters import FrechetCellFilter
a

atoms = crystal(
    symbols=['Mo', 'Se'],
    basis=[(1/3, 2/3, 1/4),
           (1/3, 2/3, 0.621)],
    spacegroup=194,
    cellpar=[3.288, 3.288, 12.900, 90, 90, 120]
    )

atoms.calc = GPAW(
    mode=PW(650),
    xc='PBE',
    kpts={'size': (8, 8, 4), 'gamma': True},
    occupations=FermiDirac(0.05),
    txt='mose2_bulk_relax.txt'
)
opt = BFGS(FrechetCellFilter(atoms), trajectory='mose2.traj', logfile='mose2.log')
opt.run(fmax=0.02)
atoms.write('wse2_relaxed.cif')
# Switch to SPR-KKR calculator
calc = SPRKKR(atoms=atoms)
print("Running SPR-KKR SCF...")
scf_result = calc.calculate()

# Reuse the same calculator and switch task — this preserves
# the converged potential from the SCF step
print("Running ARPES...")
calc.change_task('arpes')
arpes_result = calc.calculate()
print("ARPES finished.")
\end{lstlisting}

\paragraph{High-throughput composition scanning:}
The \texttt{SPRKKRAtoms} site-occupation model makes ASE2SPRKKR particularly well suited
for systematic alloy studies. By iterating over a composition parameter and updating the
fractional occupancies of a given site, one can efficiently scan the full concentration
range of a substitutional alloy series within a single script. The CPA treatment is
handled internally by SPR-KKR, so no supercells are required. The following example
scans the Fe$_x$Co$_{1-x}$ composition in a Heusler structure across 21 uniformly spaced
concentrations. The structure is read from the potential file created in the example in Section \ref{Ni2FeGa.example}:

\begin{lstlisting}[style=python]
import numpy as np
from ase.io import read
from ase2sprkkr import SPRKKR

Ni2FeGa = read("Ni2FeGa.pot")

results = []
for x in np.linspace(0, 1, 21):
    atoms = Ni2FeGa.copy()
    atoms.sites[2].site_type.occupation.set({ 'Fe': x, 'Co': 1-x })
    calc = SPRKKR(atoms=atoms)
    result = calc.calculate()
    results.append( (x, result.energies.total()) )
\end{lstlisting}

\paragraph{Multi-scale magnetic modeling:}
ASE2SPRKKR provides an interface to atomistic spin-dynamics simulations by extracting
Heisenberg exchange coupling constants $J_{ij}$ from a converged KKR potential,
exporting them in the input format expected by the UppASD code~\citep{Eriksson2017} and running the UppASD. This
two-step procedure, first-principles exchange parameters fed into a classical spin
Hamiltonian, allows finite-temperature magnetic properties such as the Curie temperature
and magnon spectra to be obtained within a single automated workflow. 
For already computed Heisenberg exchange coupling constants, utility \texttt{ase2sprkkr uppasd} can be used for creating pos, mom, and exchange files required for running UppASD ~\citep{Eriksson2017}.
\subsection{Command-Line Utilities}

The \texttt{ase2sprkkr} tool supplements the Python API:
\begin{lstlisting}[style=bash]
ase2sprkkr example            # locate and lists example scripts
ase2sprkkr shell --example 8 --jupyter # run the example 8
ase2sprkkr config -e          # edit configuration  
ase2sprkkr plot file.bsf      # visualize results
ase2sprkkr k-path sys.pot     # interactive k-path selection
ase2sprkkr uppasd JXC.out -u  # convert J_ij for UppASD
ase2sprkkr info --task=scf    # parameter documentation
\end{lstlisting}

The \texttt{k-path} tool launches an interactive Brillouin zone viewer. Clicking high-symmetry points generates SPR-KKR-compatible path specifications that automatically handle surface reciprocal lattice vectors.

\section{Applications}
\label{sec:applications}

This section demonstrates ASE2SPRKKR through representative applications that highlight different aspects of the interface. Each example is chosen to illustrate a distinct workflow pattern: task switching to spectroscopic mode (Sec.~\ref{sec:au111_arpes}), semi-infinite surface construction (Sec.~\ref{sec:au111_bsf}),  multi-scale parameter extraction with third-party export (Sec.~\ref{sec:jij}), automated basis completion for open structures (Sec.~\ref{sec:emptyspheres}), and element-selective spectroscopy (Sec.~\ref{sec:xas}). In each case, the focus is on how ASE2SPRKKR expresses the calculation as a compact Python script, replacing the manual file preparation that SPR-KKR conventionally requires.

\subsection{One-Step ARPES for Au(111)}
\label{sec:au111_arpes}

This example demonstrates task switching from a ground-state calculation to a
photoemission task on the same potential, and illustrates how
geometry-specific parameters such as surface orientation, photon energy, and
detector configuration are expressed as plain Python dictionaries. The
\texttt{arpes} task keyword invokes the full one-step photoemission
implementation \citep{Pendry1976,Braun1996}; no modification to the
calculator object or potential is required.

\subsubsection{Bulk SCF Calculation}

\begin{lstlisting}[style=python]
from ase.build import bulk
from ase2sprkkr import SPRKKR

Au = bulk('Au', 'fcc', a=4.08)
opts = {
    'CONTROL.KRMT': 4,
    'TAU.NKTAB3D': 1000,
    'SCF.VXC': 'VWN',
    'SCF.FULLPOT': False,     # ASA mode
    'SCF.NITER': 2000,
    'SCF.MIX': 0.1,
    'SCF.TOL': 1E-5,
}
calc = SPRKKR(atoms=Au, options=opts)
calc.calculate()
\end{lstlisting}

A DOS calculation on the converged potential serves as an intermediate check before launching the more expensive ARPES task.

\begin{lstlisting}[style=python]
DOS_input = {
    'ENERGY.ImE': 0.001,
    'ENERGY.EMINEV': -12.0,
    'ENERGY.EMAXEV': 7.0,
    'ENERGY.NE': 1000,
    'TAU.NKTAB': 2500,
}
result = calc.calculate(
    input_parameters='dos',
    options=DOS_input,
)
result.dos.plot()
\end{lstlisting}

\subsubsection{ARPES Intensity Calculation}

Switching to the ARPES task reuses the same potential\footnote{If the same calculator is used, the already converged potential is used automatically. Otherwise, if you created a new calculator, pass the potential filename `Au.pot_new' to the \tt{calculate} method using the \tt{potential} argument.} photon geometry,
surface orientation, and detector configuration are specified entirely within
the options dictionary:

\begin{lstlisting}[style=python]
ARPES_input = {
    'TAU.NKTAB': 500,
    'ENERGY.EMINEV': -4.0,
    'ENERGY.EMAXEV': 2.0,
    'ENERGY.EWORK_EV': 5.0,        # Work function (eV)
    'ENERGY.NE': 300,
    'ENERGY.IMV_INI_EV': 0.02,     # Initial state broadening
    'ENERGY.IMV_FIN_EV': 2.0,      # Final state broadening
    'TASK.MILLER_HKL': [1, 1, 1],  # Surface orientation
    'SPEC_PH.THETA': 45,            # Photon incident angle (deg)
    'SPEC_PH.PHI': 90,              # Photon azimuth (deg)
    'SPEC_PH.EPHOT': 50,            # Photon energy (eV)
    'SPEC_STR.N_LAYDBL': [12, 12], # Layer doubling iterations
    'SPEC_STR.N_LAYER': 30,         # Total number of layers
    'SPEC_EL.THETA': [-20, 20],    # Detector angle range (deg)
    'SPEC_EL.NT': 200,              # Angular points
    'SPEC_EL.SPOL': 4,              # Spin-resolved output
}
result = calc.calculate(
    input_parameters='arpes',
    options=ARPES_input
)
result.arpes.plot()
\end{lstlisting}

A constant-energy Fermi map is obtained by updating only the energy and
angular range, reusing all structural and photon parameters via
\texttt{dict.update()}:

\begin{lstlisting}[style=python]
ARPES_input_fermi = {
    **ARPES_input,   #copy values from the previous options
    'ENERGY.EMINEV': 0.0,
    'ENERGY.EMAXEV': 0.0,
    'ENERGY.NE': 1,
    'SPEC_EL.KA':[-0.5,-0.5], # Corner of k-grid
    'SPEC_EL.K1':[1,0], # First vector of k-grid
    'SPEC_EL.K2':[0,1], # Second vector of k-grid
    'SPEC_EL.NK1':120, # Number k-points along K1
    'SPEC_EL.NK2':120, # Number k-points along K2
    'SPEC_EL.PSPIN':[0,0,1]
}

calc.calculate(
    potential='Au.pot_new',
    input_parameters='arpes',
    options=ARPES_input_fermi
)
\end{lstlisting}

\subsection{Au(111) Surface: Bloch Spectral Function}
\label{sec:au111_bsf}

This example demonstrates ASE2SPRKKR's semi-infinite surface construction and the two-stage SCF workflow it manages automatically. Starting from a standard ASE slab, \texttt{semiinfinite\_system} builds the interaction-zone geometry; the interface then handles the sequencing of the bulk and surface SCF calculations, potential-file management, and parameter inheritance across stages.

\subsubsection{System Construction and Convergence}
\label{AuSurface.example}

\begin{lstlisting}[style=python]
from ase.build import fcc111
from ase2sprkkr.sprkkr.build import semiinfinite_system
from ase2sprkkr import SPRKKR

# Construct (111) slab: 3 atomic layers
atoms_bulk = fcc111('Au', size=(1, 1, 3), periodic=True)

# Create semi-infinite geometry: 3x3 lateral supercell, axis=2 (z)
Au111 = semiinfinite_system(
    atoms=atoms_bulk,
    repeat=[3, 3],
    atoms2=None,
    axis=2
)
\end{lstlisting}

The semi-infinite construction requires two SCF calculations. The interface automatically manages potential file sequencing between stages. First, the ``left bulk'' converges to periodic bulk properties:

\begin{lstlisting}[style=python]
from ase2sprkkr import SPRKKR
opts_bulk = {
    'CONTROL.KRMT': 2,        # Muffin-tin radius cutoff
    'SITES.NL': 4,            # Angular momentum lmax=3
    'TAU.KKRMODE': 'TB',      # Tight-binding (screened)
    'TAU.NKTAB3D': 1000,      # 3D k-mesh (~20x20x20)
    'TAU.NKTAB2D': 100,       # 2D k-mesh (~10x10)
    'TAU.CLURAD': 2.7,        # Cluster radius (Ang)
    'SCF.VXC': 'VWN',         # Vosko-Wilk-Nusair XC
    'SCF.NITER': 500,         # Number of SCF iterations
    'SCF.MIX': 0.1,           # Broyden mixing parameter
    'SCF.TOL': 1E-5,          # Potential convergence criterion
}
calc = SPRKKR(atoms=Au111, options=opts_bulk)
calc.calculate()
\end{lstlisting}

The second stage inherits the bulk options via \texttt{dict.copy()} and
overrides only the parameters that differ for the surface zone, illustrating how the interface supports incremental parameter modification without duplicating the full input:

\begin{lstlisting}[style=python]
opts_surf = opts_bulk.copy()
opts_surf.update({
    'TAU.NKTAB2D': 71,    
    'SCF.NITER': 2000,
    'SCF.MIX': 0.01,      # Smaller mixing for stability
})
calc = SPRKKR(
    potential='Au12X12.pot_new',
    options=opts_surf
)
calc.calculate()
\end{lstlisting}

\subsubsection{Bloch Spectral Function Calculation}

Bloch spectral function calculations require specifying a $k$-path along which the function is evaluated. The built-in command-line tool can assist with its generation. The example uses the converged potential file obtained in the previous example (Section \ref{AuSurface.example}).

\begin{lstlisting}[style=bash]
ase2sprkkr k-path Au12X12.pot_new_new
\end{lstlisting}

Running a spectral-function calculation requires only modifying \texttt{input_parameters}; all structural information and the potential are inherited from the previously converged calculation.

\begin{lstlisting}[style=python]
from ase2sprkkr import SPRKKR
BSF_input = {
    'TAU.NKTAB3D': 1000,
    'TAU.NKTAB2D': 71,
    'TAU.CLURAD': 2.7,
    'ENERGY.ImE': 0.002,         # 2 meV broadening
    'ENERGY.EMINEV': -4.0,
    'ENERGY.EMAXEV': 2.0,
    'ENERGY.NE': 400,
    # parameters generated by ase2sprkkr k-path
    'TASK.NK': 300,          # nb. of k-points along the path
    'TASK.KA': [[-0.66667, 0.0, 0.0], [0.0, 0.0, 0.0]], # the starting position of k-grid 
    'TASK.KE': [[0.0, 0.0, 0.0],  [0.66667, 0.0, 0.0]], # position of k-grid
}
calc = SPRKKR()
result = calc.calculate(
    potential='Au12X12.pot_new_new',
    input_parameters='bsfek',
    options=BSF_input
)
\end{lstlisting}

The output object returned by \texttt{calculate()} exposes layer-resolved
data directly, enabling post-processing without manual file parsing:

\begin{lstlisting}[style=python]
from ase2sprkkr import OutputFile
bsf = OutputFile.from_file('Au12X12_BLOCHSF_spol.bsf')
# Select surface (IQ=10,11,12) and bulk-like (IQ=0,1,2) layers
bsf.plot(IQ=0:2,10:12)
\end{lstlisting}

The same calculator instance switches to a Fermi-surface task by changing
only the \texttt{input\_parameters} keyword and the energy range:

\begin{lstlisting}[style=python]
BSF_input_fermi = {
    'TAU.NKTAB2D': 200,
    'ENERGY.EMINEV': 0.0,
    'ENERGY.EMAXEV': 0.0,
    'ENERGY.NE': 1,
    'TASK.KA': [-0.66667, -0.66667, 0.0],
    'TASK.K1': [ 0.66667, -0.66667, 0.0],
    'TASK.K2': [-0.66667,  0.66667, 0.0],
    'TASK.NK1': 200,
    'TASK.NK2': 200,
}
result = calc.calculate(
    potential='Au12X12.pot_new_new',
    input_parameters='bsfkk',
    options=BSF_input_fermi
)
\end{lstlisting}

\subsection{Exchange Coupling Parameters for Heusler Alloys}
\label{sec:jij}

This example illustrates ASE2SPRKKR's multi-scale workflow capability: exchange
parameters are extracted from the converged potential in a single additional
\texttt{calculate()} call, and the output file is then converted to the format
required by the UppASD atomistic spin-dynamics code via a built-in
command-line utility, all without leaving the Python environment or manually
editing intermediate files.

\begin{lstlisting}[style=python]
from ase2sprkkr import SPRKKR

calc = SPRKKR()
JXC_input = {
    'CONTROL.DATASET': 'Ni2FeGa',
    'TAU.NKTAB': 1000,
    'MODE.LLOYD': False,
    'TASK.CLURAD': 4.5,        # Include 3rd neighbours
    'SITES.NL': 4,
    'MODE': 'SP-SREL',         # Scalar-relativistic
    'ENERGY.NE': 32,           # Energy integration points
}
result = calc.calculate(
    potential='Ni2FeGa.pot_new',
    input_parameters='jxc',
    options=JXC_input
)
result.jxc.plot(layout = (1,3))
print(result.mean_field_curie_temperature)
result.write_uppasd_files()
\end{lstlisting}

The mean-field Curie temperature 
is computed by SPR-KKR from exchange coupling parameters as
\begin{equation}
T_C^{\text{MF}} = \frac{2}{3k_B}\sum_j J_{ij}.
\end{equation}
The method \texttt{write_uppasd_files()} converts the result to UppASD input files (\texttt{jfile}, \texttt{mfile} and \texttt{posfile}). The same can be achieved using 
commandline utility:

\begin{lstlisting}[style=bash]
ase2sprkkr uppasd Ni2FeGa_JXC.jxc
\end{lstlisting}

This bridges first-principles exchange parameters to atomistic spin dynamics
without any manual format conversion, completing a multi-scale workflow that
would otherwise require multiple specialised tools and bespoke scripts.

\subsection{Empty Sphere Generation}
\label{sec:emptyspheres}

Open covalent structures such as diamond-cubic silicon are poorly described
within the Atomic Sphere Approximation (ASA) unless the large interstitial voids are
explicitly represented. Diamond-cubic Si (space group $Fd\bar{3}m$, No.~227)
consists of an fcc Bravais lattice with a two-atom basis at $(0,0,0)$ and $(\nicefrac{1}{4},\nicefrac{1}{4},\nicefrac{1}{4})$, 

giving an atomic packing factor of only $\approx 0.34$: roughly two-thirds of the primitive cell is interstitial vacuum, making the ASA unreliable without basis completion.

ASE2SPRKKR automates this process entirely. Passing the \texttt{empty\_spheres} parameter to the \texttt{SPRKKR} constructor is sufficient; the interface identifies the symmetry-allowed interstitial sites,
assigns sphere radii under the specified overlap constraints, and writes a complete potential file without user intervention.

\begin{lstlisting}[style=python]
from ase.build import bulk
from ase2sprkkr import SPRKKR

Si = bulk('Si', 'diamond', a=5.43)

calc = SPRKKR(atoms=Si)

calc.save_input(potential_file='Si.pot', empty_spheres={
        'max_radius': 2.0,
        'min_radius': 0.2,
        'max_spheres': 300,
        'verbose': True,
    }
)
\end{lstlisting}

For this setup ASE2SPRKKR augments the two-atom primitive cell with two
empty spheres at the tetrahedral interstitial voids, yielding a four-site
basis (2~Si~+~2~ES) with equal muffin-tin and Wigner-Seitz radii
$R_{\mathrm{MT}} \approx 2.22$~a.u. and $R_{\mathrm{WS}} \approx 2.53$~a.u.
The muffin-tin filling fraction increases from $\approx 0.34$ (Si atoms alone)
to $\approx 0.68$ once the two empty spheres are included, halving the
effective interstitial region. The Wigner-Seitz radii satisfy
$\sum_i \tfrac{4\pi}{3}R_{\mathrm{WS},i}^3 = V_{\mathrm{cell}}$ 
by
construction, ensuring the conservation of the volume of the unit cell. The more homogeneous
real-space partitioning that results reduces the angular-momentum cutoff
$\ell_{\mathrm{max}}$ needed for an accurate description of the interstitial
charge density and mitigates the numerical artifacts that arise in open
tetrahedral networks when only atomic spheres are used.
In many cases, just passing \texttt{True} to the \texttt{empty_spheres} parameter of \texttt{calculate} or \texttt{save_input} methods of \texttt{SPRKKR} calculator (The parameter can also be passed directly to the calculator, in which case it will be applied to all calls of the aforementioned methods) is sufficient.
However, in some cases, better results can be achieved by fine-tuning the parameters, as illustrated in the example above. Manual addition of empty spheres using \texttt{ase2sprkkr.bindings.empty_spheres.add_empty_spheres} function is, of course, also possible.

\subsection{X-ray Absorption and Magnetic Circular Dichroism}
\label{sec:xas}

This example demonstrates element-selective spectroscopy: the \texttt{xas}
task targets a specific atomic site and core level via two keywords
(\texttt{TASK.IT} and \texttt{TASK.CL}), while the interface handles
energy-mesh generation, matrix-element integration, and output formatting.
Post-processing via the XMCD sum rules is expressed in three lines using the
structured result object.

\begin{lstlisting}[style=python]
from ase2sprkkr import SPRKKR

calc = SPRKKR()
XAS_input = {
    'CONTROL.DATASET': 'Ni2FeGa',
    'TAU.NKTAB': 500,
    'TASK.IT': 1,              # Target site: Ni
    'TASK.CL': '2p',           # Core level: 2p -> 3d
    'TASK.FRAMETET': 0,        # Polar angle
    'TASK.FRAMEPHI': 0,        # Azimuthal angle
    'ENERGY.NE': 180,
    'ENERGY.GRID': 6,
    'ENERGY.EMAX': 40.0,
}
result = calc.calculate(
    input_parameters='xas',
    options=XAS_input,
    potential='Ni2FeGa.pot_new'
)
\end{lstlisting}

Following \citep{Carra1993,Thole1992}, spin and orbital moments are extracted from the
integrated dichroic signal:
\begin{align}
\mu_L &= -\frac{4q}{3r}\,n_h\,\mu_B, \label{eq:muL} \\
\mu_S &= -\frac{6p - 4q}{r}\,n_h\,\mu_B, \label{eq:muS}
\end{align}
where $p = \int_{L_3}(I^+-I^-)\mathrm{d}E$, $q = \int_{L_3+L_2}(I^+-I^-)\mathrm{d}E$,
$r = \int_{L_3+L_2}(I^++I^-)\mathrm{d}E$, and $n_h = 10 - n_{3\mathrm{d}}$ is obtained
directly from the charge self-consistency output. The integrals $p$, $q$,
and $r$ are accessible as attributes of the result object, so the full
sum-rule analysis reduces to a few arithmetic operations on the returned data without any manual file parsing.

\section{Conclusion}
\label{sec:conclusion}

We have presented ASE2SPRKKR, a Python interface that embeds the Spin-Polarized
Relativistic Korringa-Kohn-Rostoker code within the Atomic Simulation
Environment. The central design goal is to make SPR-KKR's unique capabilities
accessible through the same scripting paradigm that ASE users already employ
for other codes: structures are built with standard ASE tools, calculations are
launched through a unified \texttt{SPRKKR} calculator, tasks are switched by
changing a single keyword, and results are returned as structured objects with
built-in plotting and export methods. The \texttt{SPRKKRAtoms} extension
accommodates fractional site occupations for CPA calculations while retaining
full ASE compatibility, and automated input validation, MPI support, and
potential file management collectively eliminate the manual bookkeeping that
has historically limited access to the code.

The five applications illustrate how this design reduces complex multi-step
workflows to compact scripts. Semi-infinite surface construction and two-stage
SCF convergence are handled by \texttt{semiinfinite\_system} and
\texttt{dict.update()} parameter inheritance. Task switching between BSF,
DOS, ARPES, JXC, and XAS calculations requires only a keyword change on the
same calculator and potential. Multi-scale export to UppASD is a single
command-line call. Automated empty-sphere placement extends the interface to
open covalent structures without user intervention. In each case the
interface, not the underlying physics code, is what changes the user
experience.

By grounding its architecture in FAIR principles of Findability,
Accessibility, Interoperability, and Reusability, ASE2SPRKKR establishes a
replicable blueprint for bringing other specialised Green's function codes
into reproducible, collaborative workflows. Future development will extend
coverage to transport property calculations via the Kubo-Bastin
formalism, workflows for correlated materials, all within the
same framework. 

\section*{Acknowledgments}
We thank the ASE development team for maintaining the open-source framework that underpins this work. 
This work was supported by the Ministry of Education, Youth and Sports of the Czech Republic through the e-INFRA CZ (ID:90254).
\subsection*{Funding}
R.E. and J.M, acknowledge funding from Horizon Europe MSCA Doctoral Network Grant No. 101073486, EUSpecLab, funded by the European Union. This work was supported by the project Quantum materials for applications in sustainable technologies (QM4ST), funded as Project No. CZ.02.01.01/00/22\_008/0004572 by Programme Johannes Amos Comenius, call Excellent Research (R.E., J.M., A.P.). M.N. acknowledges funding from the Czech Science Foundation (GAČR), project No. 23-04746S until 2025, and project No. 25-16339S from 2026 onwards.

\subsection*{Code availability}
ASE2SPRKKR is open-source software released under the MIT licence. The source code is publicly available at \url{https://github.com/ase2sprkkr/ase2sprkkr}. The SPRKKR multiple scattering package is freely available (no costs apply) under a specific user license, and it can be downloaded after registration at \url{https://sprkkr.org/software-request}
Further documentation is available at \url{https://ase2sprkkr.github.io/ase2sprkkr/}. We also encourage users to subscribe to our mailing list at \url{https://groups.google.com/g/ase2sprkkr}, where they can ask questions about the software and receive updates on new features and releases.

\subsection*{Data availability}
All input scripts required to reproduce the five workflow demonstrations in the Results section are included in the ASE2SPRKKR repository under \texttt{examples/}. The SPR-KKR potential files and raw output used to generate the figures are available from the corresponding author upon request.

\subsection*{Competing interests}
The authors declare no competing interests.

\bibliographystyle{elsarticle-num}
\bibliography{references}



\begin{figure}
    \centering
    \includegraphics[scale=0.2]{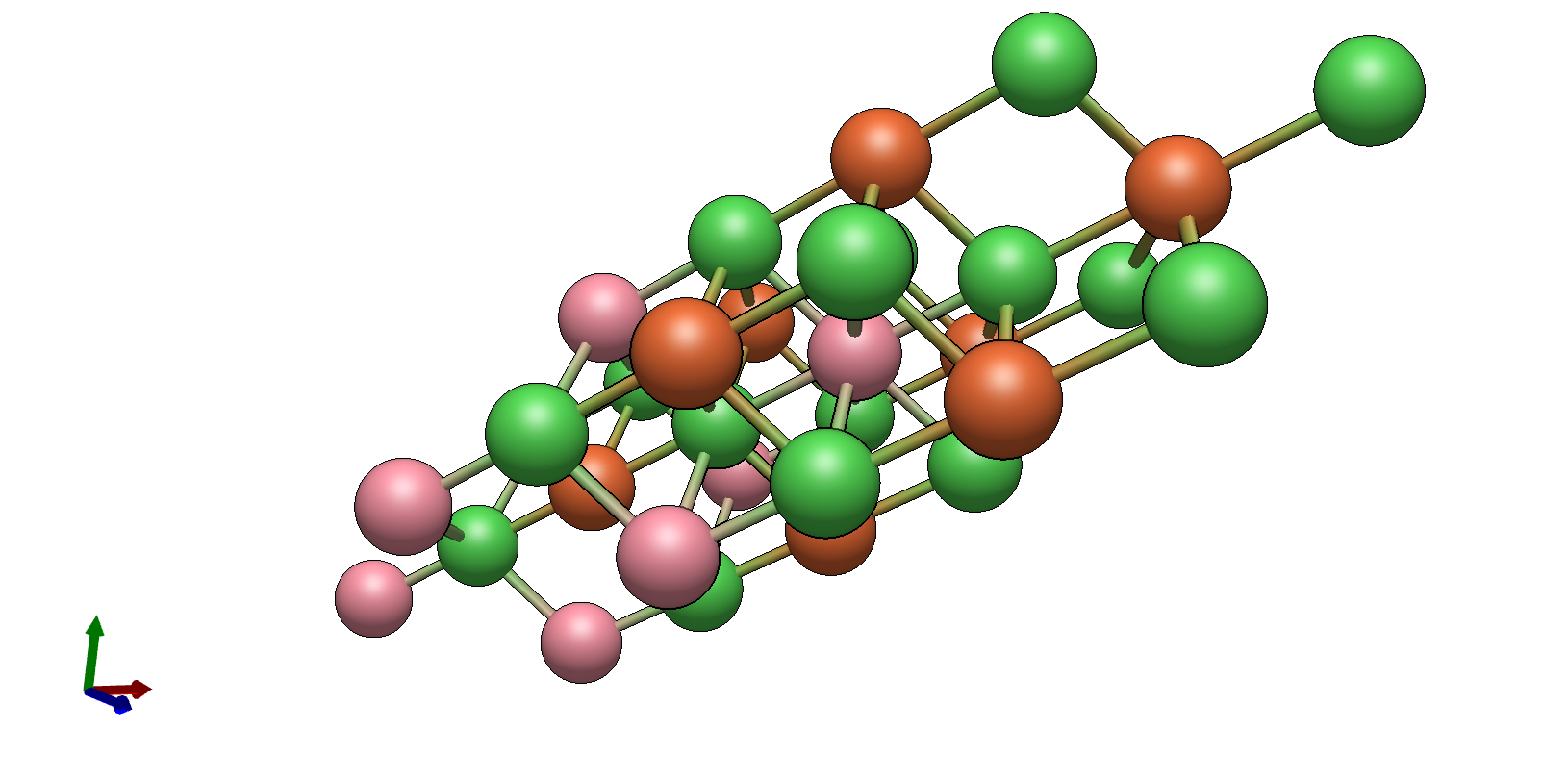}
    \caption{Full Heusler alloy Ni$_2$FeGa structure (space group F$\overline{4}$3m) with partial Fe/Co substitutional disorder. The L2$_1$ structure contains four interpenetrating fcc sublattices. Sites 2 exhibit compositional disorder with 75\% Fe / 25\% Co, representing a chemically complex magnetic Heusler compound suitable for CPA treatment.}
    \label{fig:ni2fega}
\end{figure}


\begin{figure}
\begin{subfigure}{0.9\textwidth}
\vspace{-5em}
\caption{An example of the LATTICE section of an SPR-KKR potential file}
\label{Fig:lattice:data}
\color{purple}{
\small{\begin{verbatim}
LATTICE
SYSDIM       3D
SYSTYPE      BULK
BRAVAIS            4        orthorombic primitive      mmm    D_2h
ALAT          7.1941868392
A(1)          0.0000000000   -0.5000000000   -0.5000000000
A(2)         -1.0000000000    0.0000000000    0.0000000000
A(3)          0.0000000000    1.0000000000   -1.0000000000
\end{verbatim} }}
\end{subfigure}

\hfill

\begin{subfigure}{0.9\textwidth}
\caption{The ASE2SPRKKR definition of the LATTICE section}
\label{Fig:lattice:def}
\begin{lstlisting}[style=python]
  PotSectionDefinition(
      name='LATTICE', 
      members = [
        V('SYSDIM', DefKeyword('3D')),
        V('SYSTYPE', DefKeyword('BULK')),
        V('BRAVAIS', Sequence(int, *([str]*4), allowed_values = cell_symmetries)),
        V('ALAT', float),
        V('SCALED_PRIMITIVE_CELL', Table([float]*3,
                  numbering=Integer(prefix='A(', postfix=')', format='<10'),
                  length=3L
          ),
      ],
      has_hidden_members=True,
      result_class = LatticeSection)
\end{lstlisting}
\end{subfigure}
\caption{Mapping an SPR-KKR potential file to its ASE2SPRKKR definition. (a) The raw \texttt{LATTICE} section as written in an SPR-KKR potential file. (b) The corresponding \texttt{PotSectionDefinition} in ASE2SPRKKR, which declares each member's name, expected data type, and validation rules.}

\end{figure}

\hspace{5cm}
\begin{figure}

\begin{subfigure}{0.9\textwidth}
\setlength{\abovecaptionskip}{10pt}
\setlength{\belowcaptionskip}{10pt}
\caption{The visualization of the definition of the LATTICE section (Fig.~\ref{Fig:lattice:def})}
\label{Fig:lattice:vis}
\resizebox{\textwidth}{!}{
\begin{tikzpicture}

\tikzstyle{base} = [block, rectangle, fill=blue!10]
\tikzstyle{hsection} = [fill=blue!10]
\tikzstyle{section} = [hsection, draw]
\tikzstyle{hoption} = [fill=red!10]
\tikzstyle{option} = [hoption, draw, minimum height = 0.9cm]
\tikzstyle{htype} = [fill=green!10]
\tikzstyle{type} = [htype, draw, minimum height = 0.7cm]
\tikzstyle{hhidden} = [fill=black!05]
\tikzstyle{hidden} = [hhidden, draw, minimum height = 0.7cm]

 \node [section, align = left, text depth=5.4cm, text width=12cm, minimum height=6cm, minimum width = 12cm] (lattice) {LATTICE section};   

 \node [below = of lattice.north west, anchor = north west, xshift = 2mm, yshift = 3mm, option, align = left, text width=6cm, minimum width = 4cm] (sysdim) {SYSDIM};   
 \node [right = of sysdim.west, anchor = west, xshift = 12mm, type, align = left, text width=3cm, minimum width = 3cm] (sysdimv) {Keyword('3D')};   

\node [below = of sysdim.south west, anchor = north west, yshift = 9mm, option, align = left, text width=6cm, minimum width = 4cm] (systype) {SYSTYPE}; 
 \node [right = of systype.west, anchor = west, xshift = 12mm, type, align = left, text width=3cm, minimum width = 3cm] (systypev) {Keyword('BULK')};   

\node [below = of systype.south west, anchor = north west, yshift = 9mm, option, align = left, text width=11.5cm, minimum width = 4cm, minimum height = 1cm] (bravais) {BRAVAIS}; 
 \node [right = of bravais.west, anchor = west, xshift = 12mm, type, align = left, text width=9cm, minimum width = 3cm, minimum height = 0.8cm] (bravaisv) {Sequence};   
 \node [right = of bravaisv.west, anchor = west, xshift = 11mm, type, align = left, minimum width=1.2cm, minimum height = 0.6cm] (bravaisva) {Integer};   
 \node [right = of bravaisva.west, anchor = west, xshift = 4mm, type, align = left, minimum width=1.2cm, minimum height = 0.6cm] (bravaisvb) {String};   
 \node [right = of bravaisvb.west, anchor = west, xshift = 2.8mm, type, align = left, minimum width=1.2cm, minimum height = 0.6cm] (bravaisvc) {String};   
 \node [right = of bravaisvc.west, anchor = west, xshift = 2.8mm, type, align = left, minimum width=1.2cm, minimum height = 0.6cm] (bravaisvd) {String};   
 \node [right = of bravaisvd.west, anchor = west, xshift = 2.8mm, type, align = left, minimum width=1.2cm, minimum height = 0.6cm] (bravaisve) {String};

\node [below = of bravais.south west, anchor = north west, yshift = 9mm, option, align = left, text width=6cm, minimum width = 4cm] (alat) {ALAT};   
 \node [right = of alat.west, anchor = west, xshift = 12mm, type, align = left, text width=3cm, minimum width = 3cm] (alatv) {Float};   

\node [below = of alat.south west, anchor = north west, yshift = 9mm, option, align = left, text width=11.5cm, minimum width = 4cm, minimum height = 1cm] (cell) {CELL};   
 \node [right = of cell.west, anchor = west, xshift = 12mm, type, align = left, text width=9cm, minimum width = 3cm, minimum height = 0.8cm] (cellv) {Table};   
 \node [right = of cellv.west, anchor = west, xshift = 11mm, hidden, align = left, minimum width=1.2cm, minimum height = 0.6cm] (cellva) {Integer};   
 \node [right = of cellva.west, anchor = west, xshift = 4mm, type, align = left, minimum width=1.2cm, minimum height = 0.6cm] (cellvb) {Float};   
 \node [right = of cellvb.west, anchor = west, xshift = 2.8mm, type, align = left, minimum width=1.2cm, minimum height = 0.6cm] (cellvc) {Float};   
 \node [right = of cellvc.west, anchor = west, xshift = 2.8mm, type, align = left, minimum width=1.2cm, minimum height = 0.6cm] (cellvd) {Float};   

 \node [hsection, left = of lattice.north west, anchor = north east, xshift=5mm, minimum height=1cm, minimum width=3.8cm] (sections) {Sections};   
 \node [hoption, below = of sections, minimum height=1cm, minimum width=3.8cm, yshift = 4mm] (options) {Configuration options};   
 \node [htype, yshift = 4mm, below = of options, minimum height=1cm, minimum width=3.8cm] (value) {Data types};   
 \node [hhidden, yshift = 4mm, below = of value, minimum height=1cm, minimum width=3.8cm] (help) {Hidden values};
\end{tikzpicture}
}
\end{subfigure}

\hfill

\begin{subfigure}{0.9\textwidth}
\caption{The resulting data from the example section above (Fig.~\ref{Fig:lattice:data})}
\label{Fig:lattice:python}
\small
\begin{lstlisting}[style=python]
   LatticeSection<{
       'SYSDIM': '3D',
       'SYSTYPE': 'BULK'
       'BRAVAIS': [4, 'orthorombic', ' primitive', 'mmm', 'D_2h'],
       'ALAT': 7.1941868392,
       'CELL': array([[ 0. , -0.5, -0.5],
                      [-1. ,  0. ,  0. ],
                      [ 0. ,  1. , -1. ]]),
   }>
\end{lstlisting}
\end{subfigure}
\caption{From definition to parsed object: how ASE2SPRKKR handles the LATTICE section. (a) A schematic of the definition in Fig.~\ref{Fig:lattice:def}, showing the nested hierarchy of section, configuration options, data types, and hidden values. (b) The resulting LatticeSection object obtained by parsing the data in Fig.~\ref{Fig:lattice:data}, with each entry converted to its native Python type.}
\label{Fig:lattice}
\end{figure}


\begin{figure}
    \centering
    \includegraphics[width=0.9\linewidth]{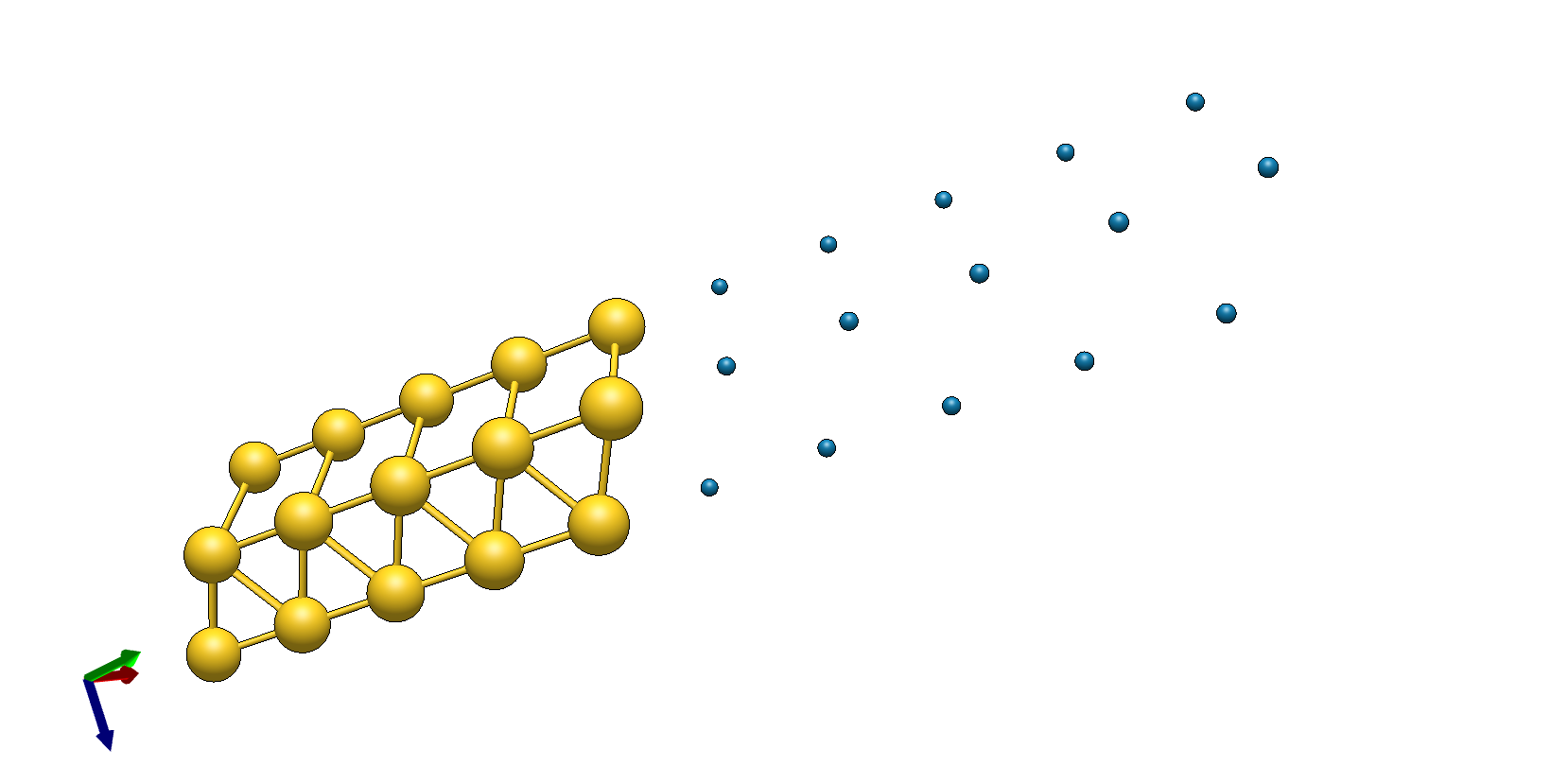}
    \caption{Semi-infinite Au(111) interaction zone constructed by
    \texttt{semiinfinite\_system}: $3{\times}3$ supercell with 12 atomic
    layers. Spheres indicate the \texttt{CLURAD}=2.7~\AA\ tight-binding region
    where inter-site scattering is treated explicitly.}
    \label{fig:au111_interaction}
\end{figure}


\begin{figure}
    \centering
    \includegraphics[width=0.5\linewidth]{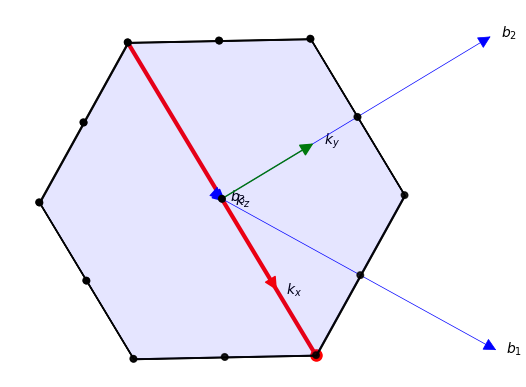}
    \caption{Interactive Brillouin zone tool launched by
    \texttt{ase2sprkkr k-path}. The hexagonal zone shows high-symmetry points
    $\bar{\Gamma}$, $\bar{M}$, and $\bar{K}$; clicking defines the path and
    the tool writes the corresponding SPR-KKR coordinate blocks directly into
    the input dictionary.}
    \label{fig:bz_gui}
\end{figure}


\begin{figure}
    \centering
    \includegraphics[width=\linewidth]{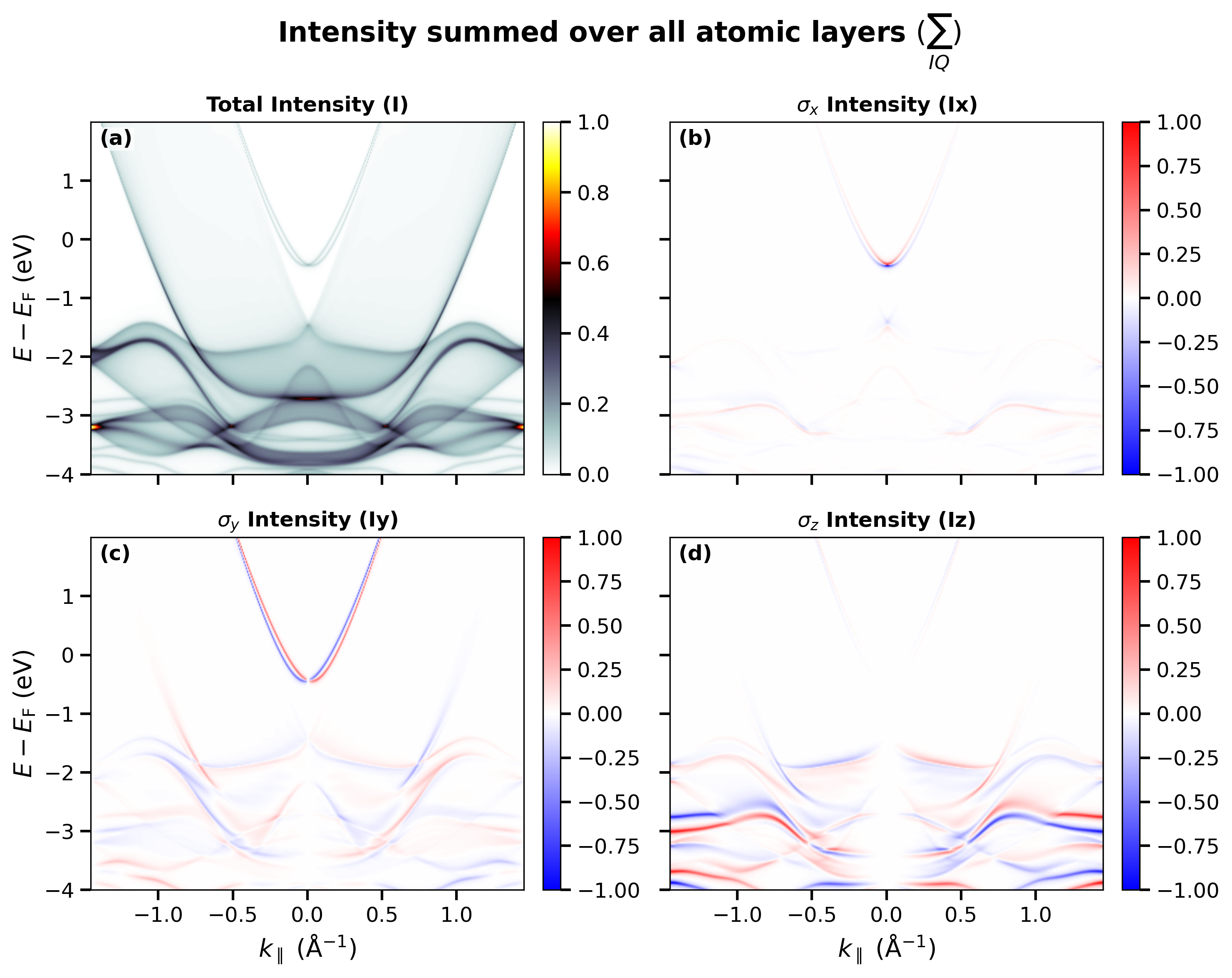}
    \caption{Spin-resolved Bloch spectral function $A(k_\parallel, E)$ for Au(111) along $\bar{K}$-$\bar{\Gamma}$--$\bar{K}$, with intensity summed over all atomic layers ($\sum_{IQ}$), produced by the \texttt{bsfek} task. (a) Total intensity $I$, showing the bulk continuum and the Rashba-split surface state crossing $E_F$ near $\bar{\Gamma}$. (b-d) The three Cartesian spin components $\sigma_x$, $\sigma_y$, $\sigma_z$ ($I_x$, $I_y$, $I_z$), where red and blue denote opposite spin polarization.}
    \label{fig:bsf_ek}
\end{figure}


\begin{figure}
    \centering
    \includegraphics[width=0.8\linewidth]{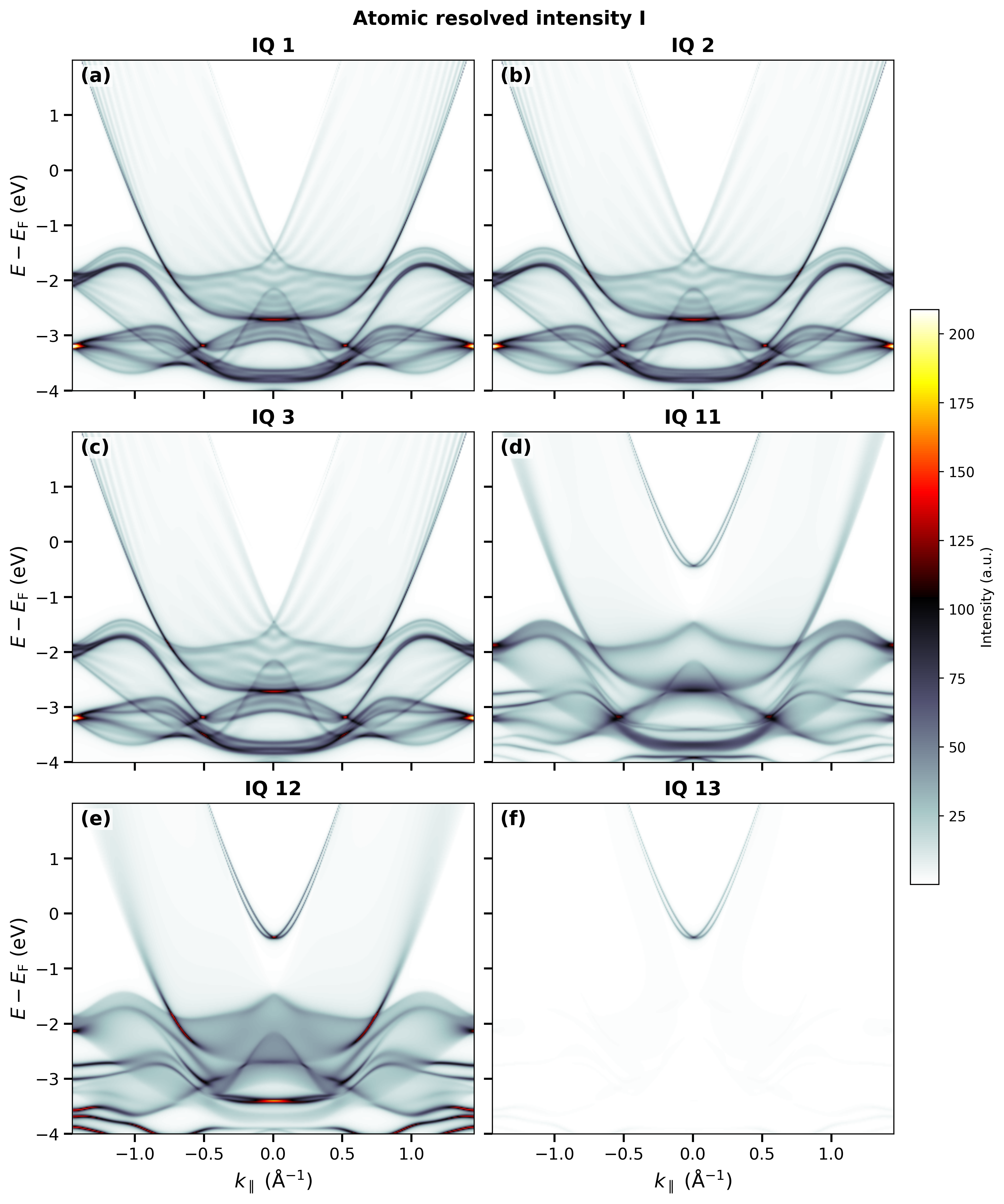}
    \caption{ Layer-resolved spectral function of Au(111), with each panel showing the Bloch spectral function projected onto a single atomic layer (IQ). Deep bulk layers (IQ 1–3, panels a–c) are essentially indistinguishable, reproducing a uniform sp continuum and confirming convergence of the bulk electronic structure. The subsurface region (IQ 11, panel d) shows the emergence of layer-localized states as bulk weight is lost. The topmost layers (IQ 12–13, panels e–f) are dominated by the surface, with the Shockley surface state near $\Gamma$ carrying most of its intensity in the outermost layer. Energy is referenced to the Fermi level $E_{F}$; intensity (colour bar, right) is in arbitrary units.}
    \label{fig:bsf_layer}
\end{figure}


\begin{figure}
    \centering
    \includegraphics[width=1\linewidth]{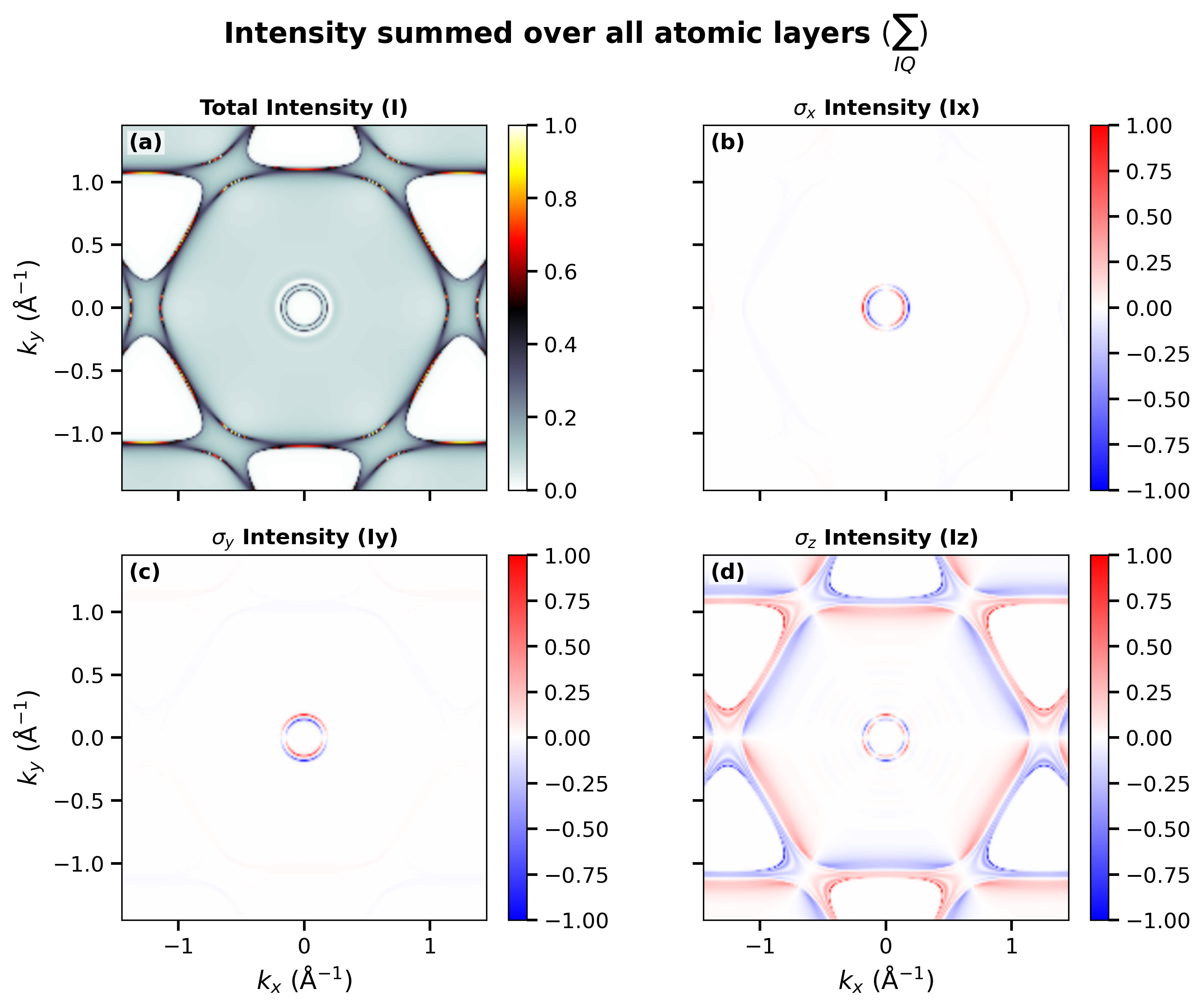}
    \caption{Spin-resolved Fermi surface $A(k_x, k_y, E_F)$ for Au(111), with intensity summed over all atomic layers ($\sum_{IQ}$), obtained from the \texttt{bsfek} task. (a) Total intensity $I$, showing the bulk Fermi contours together with the concentric Rashba-split rings of the surface state at $\bar{\Gamma}$. (b-d) The three Cartesian spin components $\sigma_x$, $\sigma_y$, $\sigma_z$ ($I_x$, $I_y$, $I_z$), where red and blue denote opposite spin polarization.}
    \label{fig:bsf_fermi}
\end{figure}


\begin{figure}
    \centering
    \includegraphics[width=0.8\linewidth]{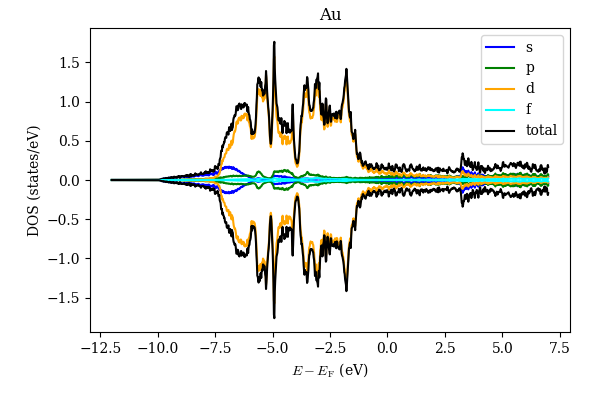}
    \caption{Bulk Au DOS obtained via the \texttt{dos} task on the
    self-consistent potential. The result object returned by
    \texttt{calculate()} is passed to a built-in \texttt{plot()} method,
    requiring no external post-processing.}
    \label{fig:dos_au}
\end{figure}


\begin{figure}
    \centering
    \includegraphics[width=1\linewidth]{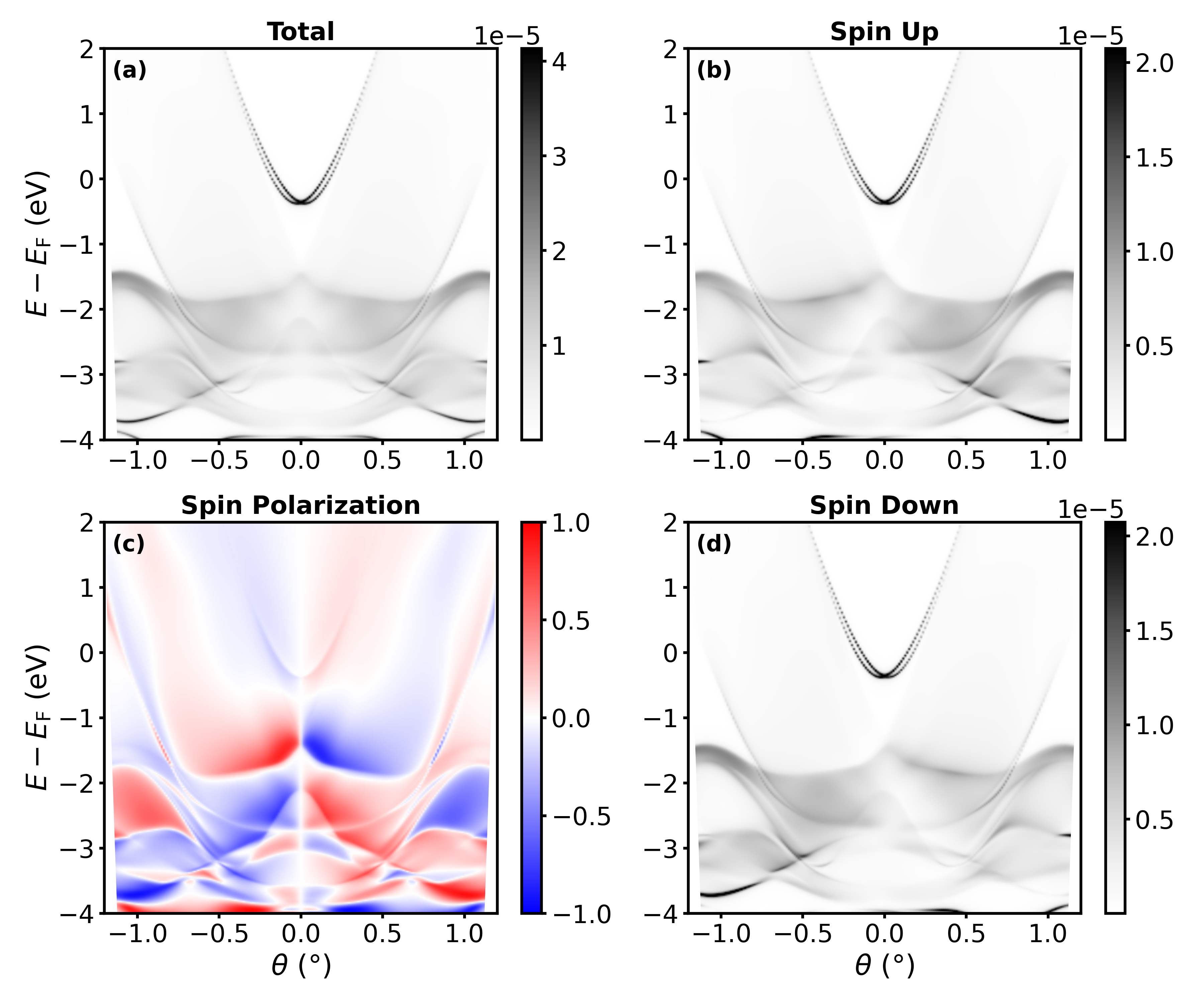}
     \caption{ One-step ARPES $I(\theta, E)$ at $h\nu$ = 50 eV returned by the \texttt{arpes} task: (a) total intensity, (b) spin-up, (c) spin polarization, and (d) spin-down. Rashba spin--momentum locking is visible as a spin-resolved intensity asymmetry, $P_{\mathrm{spin}}(\theta) = [I_\uparrow(\theta) - I_\downarrow(\theta)]/[I_\uparrow(\theta) + I_\downarrow(\theta)]$. All four panels are produced in a single calculation and accessed through the structured result object.}
    \label{fig:arpes_ek}
\end{figure}


\begin{figure}
    \centering
    \includegraphics[width=1\linewidth]{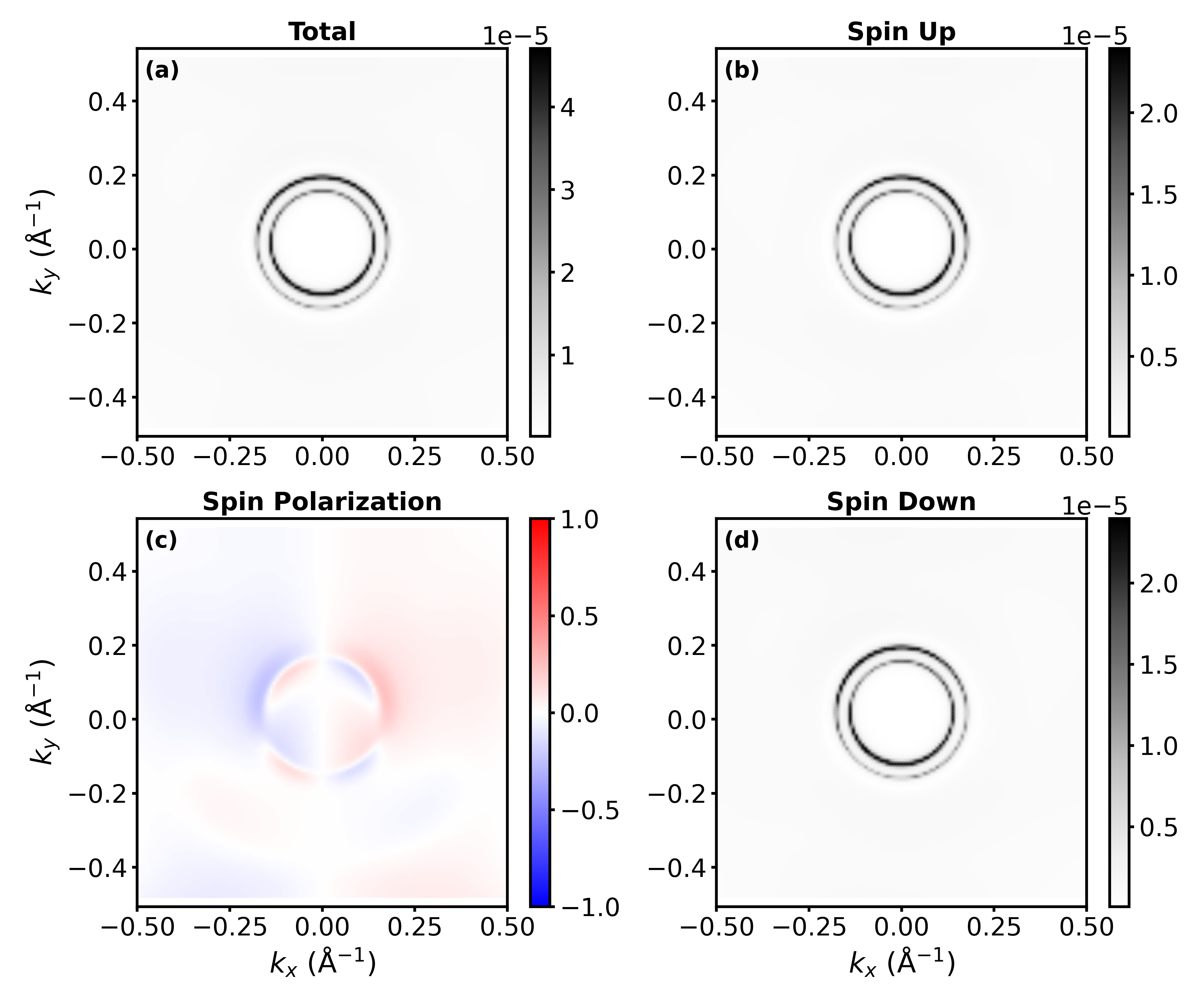}
    \caption{ One-step ARPES Fermi map $I(k_x, k_y, E_F)$ at h$\nu$ = 50 eV around the $\bar{\Gamma}$ point for Au(111), returned by the \texttt{arpes} task: (a) total intensity, (b) spin-up, (c) spin polarization, and (d) spin-down. The Rashba-split surface state appears as two concentric rings, and the spin--momentum locking is visible as a spin-resolved intensity asymmetry, $P_{\mathrm{spin}}(k) = [I_\uparrow(k) - I_\downarrow(k)]/[I_\uparrow(k) + I_\downarrow(k)]$, which reverses sign between the inner and outer ring and winds tangentially around $\bar{\Gamma}$. All four panels are produced in a single calculation and accessed through the structured result object.}    \label{fig:arpes_fermi}
\end{figure}


\begin{figure}
    \centering
    \includegraphics[width=0.8\linewidth]{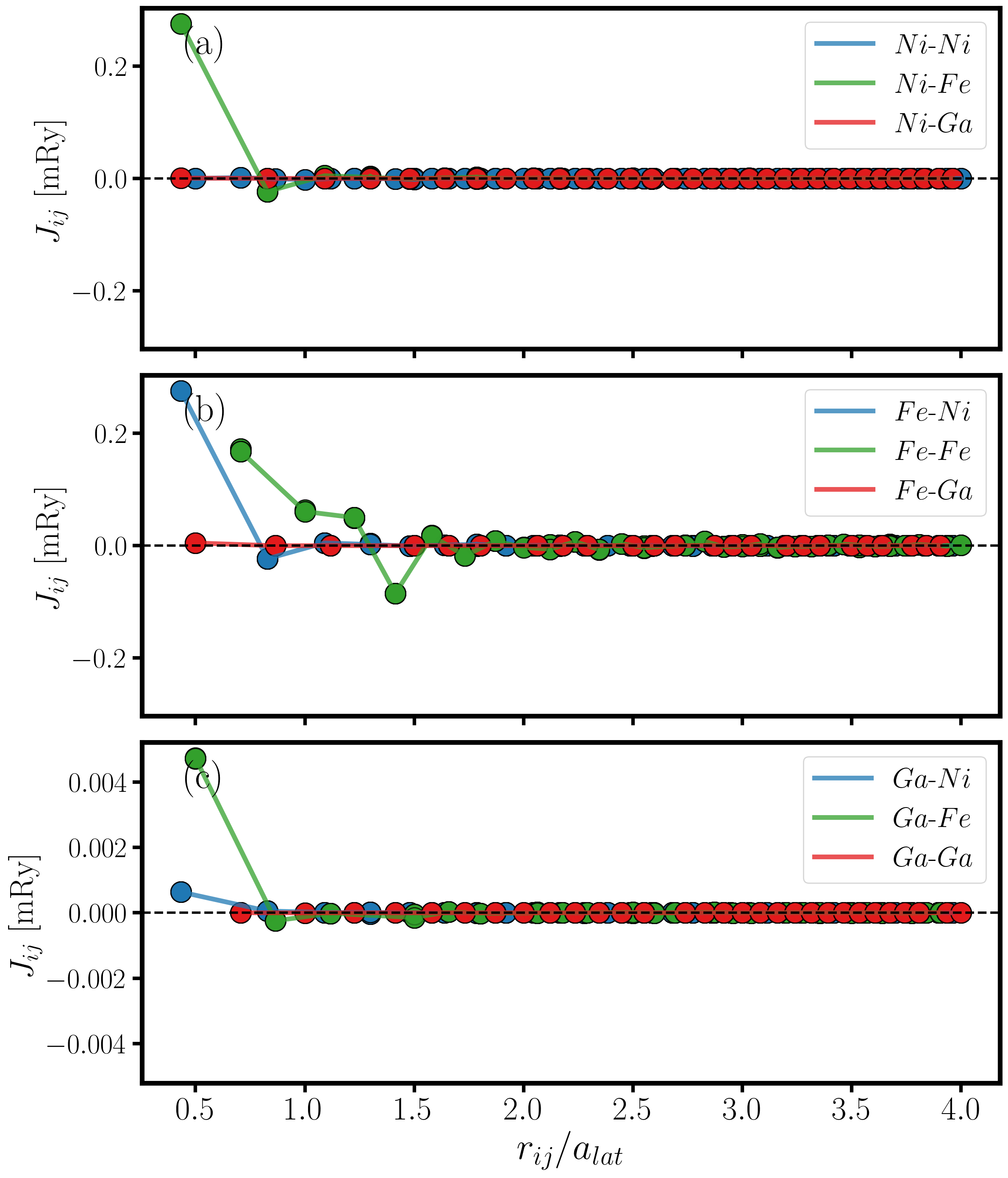}
    \caption{Exchange couplings $J_{ij}(R)$ for Ni$_2$FeGa produced by the
    \texttt{jxc} task: (a) Ni, (b) Fe, (c) Ga sublattice contributions as a
    function of inter-site distance. The structured result object provides
    per-sublattice access without post-processing of the raw output file.}
    \label{fig:jij}
\end{figure}


\begin{figure}
    \centering
    \includegraphics[width=0.8\linewidth]{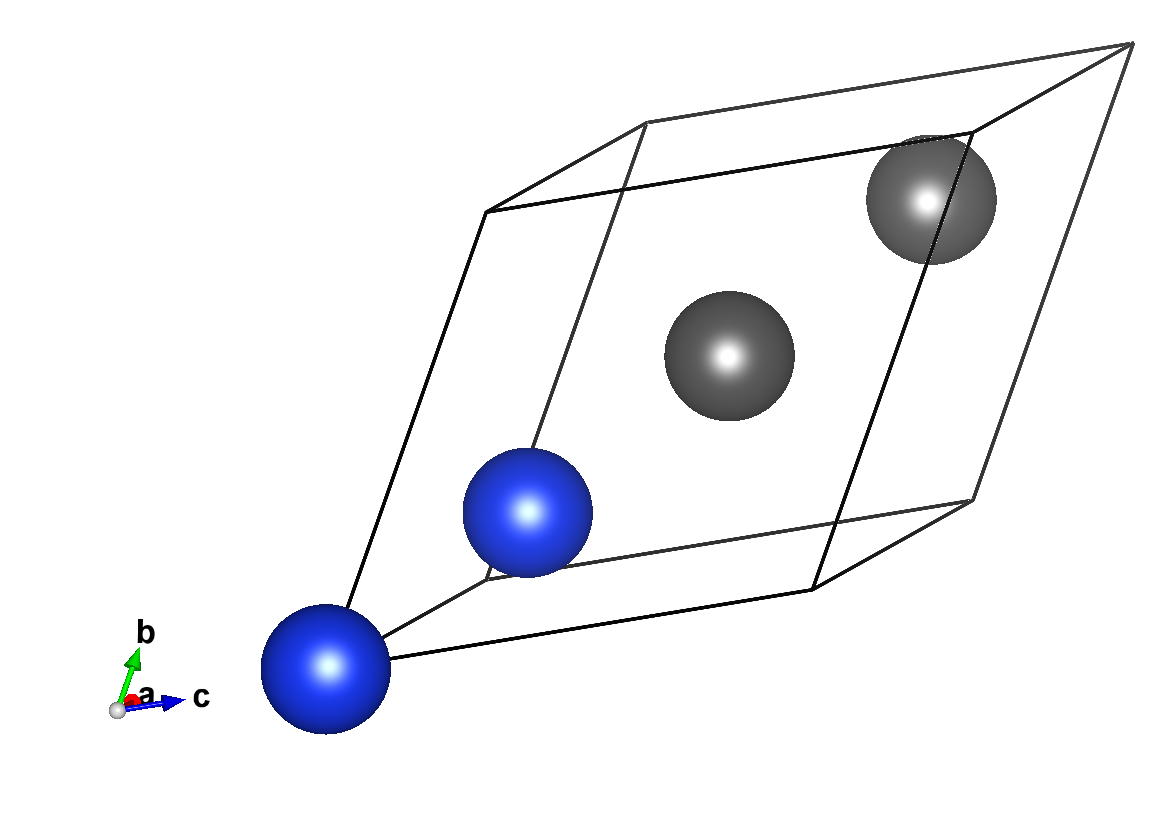}
    \caption{Diamond-cubic Si in the primitive fcc representation used by
    SPR-KKR. Blue spheres: Si atoms at $(0,0,0)$ and
    $(\tfrac{1}{4},\tfrac{1}{4},\tfrac{1}{4})$. Grey spheres: empty spheres
    placed automatically at the tetrahedral interstitial voids.
    The muffin-tin filling fraction increases from $\approx 0.34$ (Si only)
    to $\approx 0.68$ upon inclusion of the empty spheres.}
    \label{fig:si_empty}
\end{figure}


\begin{figure}
    \centering
    \includegraphics[width=1\linewidth]{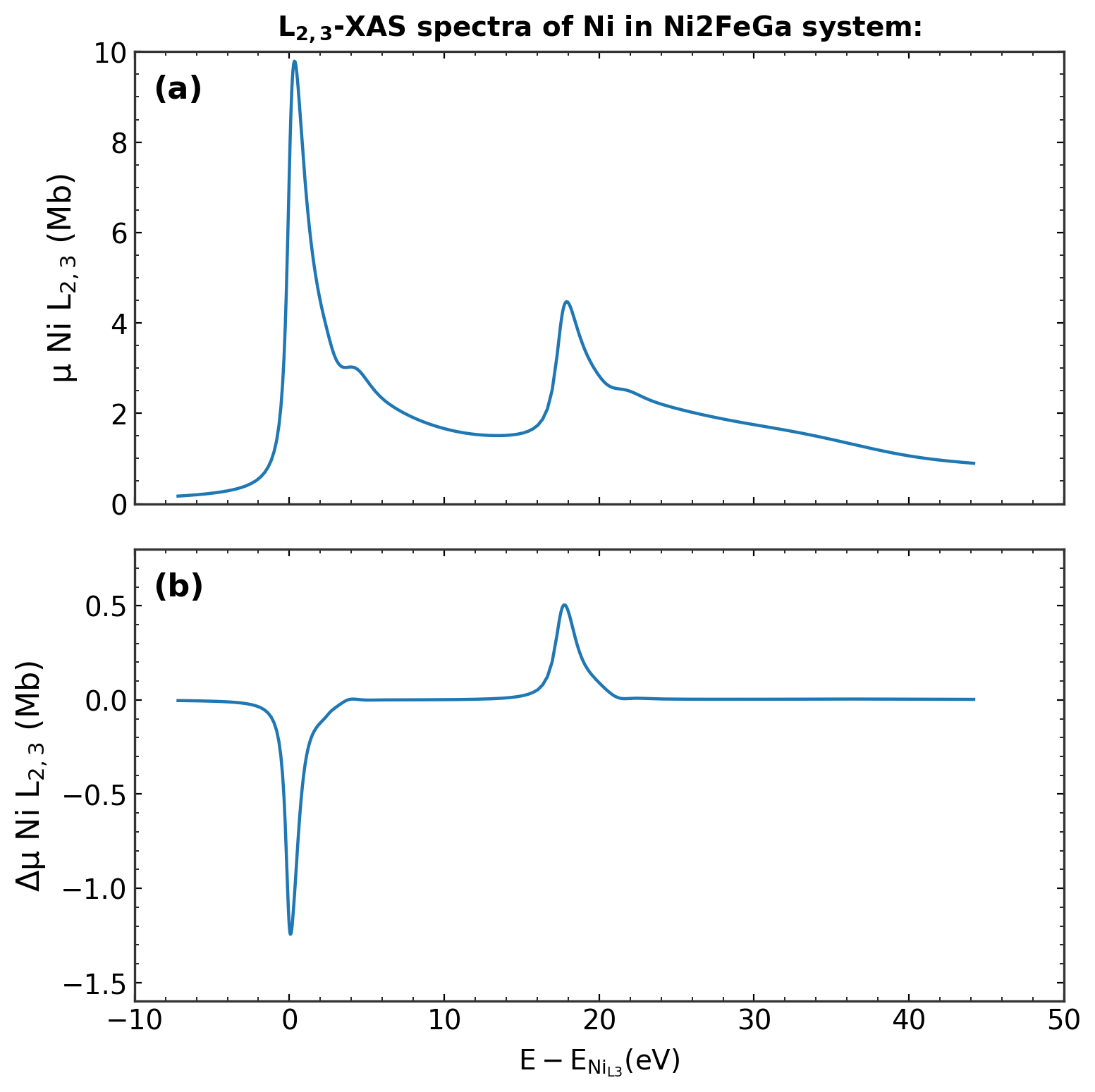}
    \caption{Ni $L_{2,3}$ XAS (a) and XMCD (b) for Ni$_2$FeGa obtained
    via the \texttt{xas} task. Targeting a specific site and edge requires only
    setting \texttt{TASK.IT} and \texttt{TASK.CL}; the interface manages
    energy integration and outputs both spectra in the result object.}
    \label{fig:xas}
\end{figure}

\end{document}


\maketitle

\begin{center}
{\small
$^{1}$ New Technologies-Research Centre, University of West Bohemia in Pilsen, 30100 Pilsen, Czech Republic \\
$^{2}$ FZU – Institute of Physics, Czech Academy of Sciences, v.v.i, Prague, Czech Republic \\
$^{3}$ Department Chemie, Ludwig-Maximilians-Universität München, Butenandtstr. 5-11, 81377 München, Germany
}
\end{center}

\section*{Theoretical Background: The KKR-Green's Function Method}
\label{sec:theory}
This section provides essential background on the KKR-Green's function formalism implemented in SPR-KKR, focusing on aspects directly relevant to the ASE2SPRKKR interface design. 
\section{Fundamental Formulation}
\label{sec:fundamental}

SPR-KKR solves the fully relativistic Dirac-Kohn-Sham equation  \citep{Ebert2011, MacDonald1979, Rose1961}
\begin{equation*}
\left[-\mathrm{i}\hbar c\,\vec{\alpha}\cdot\vec{\nabla}
      +\vec{\beta} m c^{2}
      +V_{\mathrm{eff}}(\mathbf{r})
      +\vec{\beta}\,\vec{\sigma}\cdot\mathbf{B}(\mathbf{r})
      +\vec{\alpha}\cdot\mathbf{A}(\mathbf{r})\right]
\psi(\mathbf{r}) = E\,\psi(\mathbf{r}),
\label{eq:dirac}
\end{equation*}

where $\vec{\alpha}$ and $\vec{\beta}$ are $4\times4$ Dirac matrices and $\psi$ is a four-component spinor. The effective potential $V_{\mathrm{eff}}$ incorporates the Kohn-Sham potential and exchange-correlation contributions. The magnetic field
$\vec{B}(\vec{r})=\vec{B}_{\mathrm{ext}}(\vec{r})+\partial E_{\mathrm{xc}}[n,\vec{m}]/\partial\vec{m}(\vec{r})$ combines any external field with the spin-dependent exchange-correlation field, while the vector potential $\vec{A}$ couples to the electronic current density $\vec{j}=-|e|c\,\vec{\alpha}$.

The multiple scattering formulation \citep{Gonis2000, Ebert2011}  partitions space into non-overlapping atomic regions. Single-site Dirac equations are solved for each atom type, yielding $t$-matrices $t^{n}_{\Lambda\Lambda'}(E)$ with combined quantum numbers $\Lambda=(\kappa,\mu)$, where $\kappa$ and $\mu$ are the relativistic spin-orbit and magnetic quantum numbers, respectively.  The $\Lambda$ sums are truncated at a maximum angular momentum $\ell_{\max}$, which is an important convergence parameter exposed in ASE2SPRKKR (see Section 2).  The structural Green's function $G^{0}(E)$ describes free-electron propagation between sites and depends only on the geometry. These ingredients are combined through the scattering path operator:
\begin{equation}
\underline{\tau}^{nn'}(E)
= \frac{1}{\Omega_{\mathrm{BZ}}}
  \int_{\Omega_{\mathrm{BZ}}}\mathrm{d}^{3}k\,
  \left[\underline{t}(E)^{-1}-\underline{G}^{0}(\vec{k},E)\right]^{-1}
  \mathrm{e}^{\mathrm{i}\vec{k}\cdot(\vec{R}_{n}-\vec{R}_{n'})},
\label{eq:tau}
\end{equation}
from which the site-resolved Green's function follows:
\begin{align}
G(\vec{r},\vec{r}',E)
&= \sum_{\Lambda\Lambda'}
   Z_{\Lambda}^{n}(\vec{r},E)\,
   \tau_{\Lambda\Lambda'}^{nn'}(E)\,
   Z_{\Lambda'}^{n'\times}(\vec{r}',E)
\notag\\
&\quad -\delta_{nn'}\sum_{\Lambda}
   \Big[Z_{\Lambda}^{n}(\vec{r},E)\,J_{\Lambda}^{n\times}(\vec{r}',E)\,\Theta(r'-r)
\notag\\
&\qquad\qquad +\;J_{\Lambda}^{n}(\vec{r},E)\,Z_{\Lambda}^{n\times}(\vec{r}',E)\,\Theta(r-r')\Big],
\label{eq:green}
\end{align}
where $\vec{r}$ ($\vec{r}'$) are cell-centred coordinates in atomic cell $n$ ($n'$), the superscript $\times$ denotes the left-hand-side solution, and $Z_{\Lambda}^{n}$, $J_{\Lambda}^{n}$ are the regular and irregular single-site solutions.

With the Green's function available, all ground-state observables follow directly. The density of states and electron density are given by
\begin{equation}
n(E)=-\frac{1}{\pi}\operatorname{Im}\operatorname{Tr}
      \int_{\Omega_{n}}\mathrm{d}^{3}r\;G(\vec{r},\vec{r},E),
\label{eq:dos}
\end{equation}
\begin{equation}
\rho(\vec{r})=-\frac{1}{\pi}\operatorname{Im}\operatorname{Tr}
              \int^{E_{\mathrm{F}}}\mathrm{d}E\;G(\vec{r},\vec{r},E),
\label{eq:density}
\end{equation}
where the trace applies when $G$ is given in matrix form. In practice, the energy integrals in Eqs.~(\ref{eq:density}) and~(\ref{eq:density_matrix}) are performed along a contour in the complex energy plane rather than on the real axis, exploiting the analyticity of $G$ to achieve rapid convergence with far fewer energy points than real-axis integration would require. The contour parameters, start energy, number of points, and imaginary part, are exposed as convergence parameters in ASE2SPRKKR. Notably, ground-state quantities require only site-diagonal scattering path operators $\underline{\tau}^{nn}$.

For a $\vec{k}$-resolved representation of the electronic structure, the Bloch spectral function \citep{Gyorffy1985BSF, Faulkner1980} is defined as
\begin{equation}
A_{\mathrm{B}}(\vec{k},E)
= -\frac{1}{\pi N}\operatorname{Im}\operatorname{Tr}
  \sum_{n,n'}^{N}\mathrm{e}^{\mathrm{i}\vec{k}\cdot(\vec{R}_{n}-\vec{R}_{n'})}
  \int_{\Omega}\mathrm{d}^{3}r\;
  G(\vec{r}+\vec{R}_{n},\vec{r}+\vec{R}_{n'},E).
\label{eq:BSF}
\end{equation}
For a perfectly ordered system, $A_{\mathrm{B}}(\vec{k},E)$ reduces to a sum of $\delta$-functions $\delta(E-E_{\vec{k}})$, recovering the usual band structure. For disordered systems, it provides a broadened but physically meaningful spectral representation.
\section{Coherent Potential Approximation}
\label{sec:cpa}

A key advantage of the KKR method is its natural connection to the coherent potential approximation (CPA) for substitutionally disordered alloys~\citep{Soven1967, Taylor1967, Gyorffy1972}.. For a binary alloy with components A and B at concentrations $x_A$ and $x_B$, the CPA effective medium is determined by the condition that embedding either component produces no additional scattering on average:
\begin{equation}
x_A\,\underline{\tau}_A^{nn} + x_B\,\underline{\tau}_B^{nn}
= \underline{\tau}_{\mathrm{CPA}}^{nn},
\label{eq:cpa_condition}
\end{equation}
with the component-projected scattering path operator
\begin{equation}
\underline{\tau}_{A}^{nn}
= \underline{\tau}_{\mathrm{CPA}}^{nn}
  \left[1+\left(t_{A}^{-1}-t_{\mathrm{CPA}}^{-1}\right)
  \tau_{\mathrm{CPA}}^{nn}\right]^{-1}.
\label{eq:cpa_component}
\end{equation}
The computational cost is independent of composition, making dilute impurities and equiatomic alloys equally tractable. SPR-KKR extends the CPA to multi-component alloys \citep{Faulkner1980, Turek1997} and non-collinear magnetic structures. In ASE2SPRKKR, disorder is specified through fractional site occupations in the \texttt{SPRKKRAtoms} structure, which automatically generates the appropriate CPA input.

The same CPA framework naturally accommodates the disordered local moment (DLM) approach \citep{Gyorffy1985, Staunton2014}, which maps finite-temperature magnetic fluctuations onto orientational disorder. In the paramagnetic state, each site carries a local moment whose orientation is treated as a random variable, with spin-up and spin-down orientations entering the CPA on equal footing. This enables \textit{ab initio} predictions of paramagnetic electronic structure, Curie temperatures, and magnon spectra without requiring supercells or empirical magnetic models.

\section{Spectroscopic Observables}
\label{sec:spectroscopy}

The Green's function formulation provides a unified framework for computing spectroscopic response functions, including transition matrix elements and final-state effects, which are essential for quantitative comparison with experiment.

\subsection{Valence band photoemission}

Angle-resolved photoemission spectroscopy (ARPES) \citep{Hufner2003, Damascelli2003, Lv2019} is described within the one-step model, which expresses the photocurrent as \citep{Pendry1976, Braun1996, Braun2006, Minar2020}
\begin{equation}
j_{km_s}^{q,\lambda}(E_{\mathrm{f}})
\propto \operatorname{Im}\int \mathrm{d}^{3}r\int \mathrm{d}^{3}r'\;
\left[T\phi_{km_s}^{\mathrm{LEED}}(\vec{r},E_{\mathrm{f}})\right]^{\dagger}
X_{\vec{q}\lambda}(\vec{r})\,
G(\vec{r},\vec{r}',E_i)\,
X_{\vec{q}\lambda}^{\dagger}(\vec{r}')\,
\phi_{km_s}^{\mathrm{LEED}}(\vec{r}',E_{\mathrm{f}}),
\label{eq:arpes}
\end{equation}
where $G(E_i)$ represents the initial valence-band states, $\phi^{\mathrm{LEED}}$ is the time-reversed low-energy electron diffraction final state at $E_f = E_i + \hbar\omega$, $X_{\vec{q}\lambda}$ is the electron-photon interaction operator, and $T$ is the time-reversal operator. This formulation naturally incorporates matrix-element effects  \citep{Fadley1992, Schattke2003}, surface contributions, and finite-lifetime broadening through a self-energy $\Sigma(E)$, enabling quantitative modelling of spin-orbit-induced Rashba splittings at surfaces~\citep{LaShell1996},  enabling quantitative modelling of spin-orbit-induced Rashba splittings at surfaces \citep{LaShell1996, Rashba1960}.

\subsection{Core level spectroscopies}

X-ray absorption spectroscopy (XAS) probes element-specific electronic structure. Using the Green's function identity$-\pi^{-1}\operatorname{Im}G^{+}(E)=\sum_{\alpha}|\Psi_{\alpha}\rangle\langle\Psi_{\alpha}|\,\delta(E_{\alpha}-E)$, the absorption coefficient for radiation with wave vector $\vec{q}$ and polarization $\lambda$ takes the form \citep{Ankudinov1998, Ebert1996, Arola1997}:
\begin{align}
\mu^{\bar{q}\lambda}(\omega)
&\propto \sum_{i\,\mathrm{occ}}
\langle \Phi_i | X_{\bar{q}\lambda}^{\times}\,
\Im\, G^{+}(E_i+\hbar\omega)\, X_{\bar{q}\lambda} | \Phi_i \rangle  \times \theta(E_i+\hbar\omega - E_{\mathrm{F}}),
\label{eq:85}
\end{align}

with the operator $X_{\bar{q}\lambda} = -\frac{e}{c}\,\vec{j}_{\mathrm{el}}\cdot\vec{A}_{\bar{q}\lambda}$.
This expression covers both the XANES ($0$-$50$~eV above the edge) and EXAFS  ($>50$~eV)  regimes. For magnetically ordered systems with polarized X-rays, the formalism yields X-ray magnetic circular dichroism (XMCD)\citep{Stohr1999, Ebert1996}, from which spin and orbital moments can be extracted via the XMCD sum rules~\citep{Thole1992,Carra1993}.

\section{Exchange coupling constants}
\label{sec:exchange}

Magnetic exchange coupling constants $J_{ij}$ between atomic sites $i$ and $j$ are obtained from the Liechtenstein formula~\citep{Liechtenstein1987}, which maps the energy change upon relative rotation of magnetic moments onto a classical Heisenberg Hamiltonian $H = -\sum_{ij}J_{ij}\,\hat{\mathbf{m}}_i\cdot\hat{\mathbf{m}}_j$:
\begin{equation}
J_{ij}
= -\frac{1}{4\pi}\int^{E_{\mathrm{F}}}\mathrm{d}E\;
  \operatorname{Tr}\left[
  (t_i^{\uparrow -1}-t_i^{\downarrow -1})\,\tau_{ij}^{\uparrow}\,
  (t_j^{\uparrow -1}-t_j^{\downarrow -1})\,\tau_{ji}^{\downarrow}
  \right].
\label{eq:Jij}
\end{equation}
In the fully relativistic implementation of SPR-KKR, spin-orbit coupling is incorporated at all orders and the formula is generalized to treat non-collinear configurations and anisotropic exchange \citep{Udvardi2003, Ebert2009}. These parameters serve as input for subsequent finite-temperature studies via mean-field theory, Monte Carlo simulations, or spin-dynamics approaches \citep{Skubic2008, Evans2014, Eriksson2017, Bergman2010}, enabling realistic modelling of Curie temperatures and spin-wave spectra.

\section{Transport Properties via the Kubo-Greenwood Formalism}
\label{sec:transport}

The electrical conductivity at $T=0$~K is obtained from the Kubo-Greenwood equation~\citep{Kubo1957,Greenwood1958,Butler1985}:
\begin{equation}
\sigma_{\mu\nu}
= \frac{\hbar}{\pi N\Omega}\operatorname{Tr}
  \left\langle J_{\mu}\,\operatorname{Im}G^{+}(E_{\mathrm{F}})\,
              J_{\nu}\,\operatorname{Im}G^{+}(E_{\mathrm{F}})\right\rangle_c,
\label{eq:kubo}
\end{equation}
where the subscript $c$ denotes a configurational average for disordered systems. Within the KKR-CPA framework~\citep{Butler1985}, this decomposes into $\sigma_{\mu\nu}=\sigma_{\mu\nu}^{0}+\sigma_{\mu\nu}^{1}$, where $\sigma^{0}$ and $\sigma^{1}$ are site-diagonal and off-diagonal contributions involving the component-projected scattering path operators and current-density matrix elements $j_{\mu,\Lambda\Lambda'}^{i}$. The off-diagonal term contains vertex corrections that account for the difference between correlated and uncorrelated configurational averages, analogous to the scattering-in terms of the Boltzmann formalism. These corrections are essential for dilute alloys but become negligible for concentrated alloys, consistent with Boltzmann-level treatments.

\section{Exchange-Correlation Treatment and Correlation Extensions}
\label{sec:xc}

For systems requiring treatment of strong correlations beyond standard LSDA or
GGA, SPR-KKR supports both the LDA$+U$ and the dynamical mean-field theory
(DMFT) frameworks \citep{Liechtenstein1995, Anisimov1997, Georges1996, Kotliar2006,
Minar2011, Minar2005}. Both approaches share a common starting point: the
orbital-resolved density matrix $n_{\Lambda\Lambda'}^{n}$ constructed by projecting
the Green's function onto a set of localized reference states $\Phi_{\Lambda}(\mathbf{r})$
of appropriate $d$- or $f$-character, determined at a reference energy $E_{\Lambda}$:
\begin{align}
n_{\Lambda\Lambda'}^{n}
&=
-\frac{1}{\pi}
\int^{E_F} \mathrm{d}E
\sum_{\Lambda''\Lambda'''}
\int_{\Omega_n}\mathrm{d}^3r\;
Z_{\Lambda''}^{n}(\mathbf{r},E)\,\Phi_{\Lambda}(\mathbf{r})\,
\tau_{\Lambda''\Lambda'''}^{nn}(E)
\notag\\
&\quad\times
\int_{\Omega_n}\mathrm{d}^3r'\;
\Phi_{\Lambda'}(\mathbf{r}')\,Z_{\Lambda'''}^{n}(\mathbf{r}',E)
\notag\\
&\quad-
\sum_{\Lambda''}
\int_{\Omega_n}\mathrm{d}^3r\;
Z_{\Lambda''}^{n}(\mathbf{r},E_F)\,\Phi_{\Lambda}(\mathbf{r})\,
\int_{\Omega_n}\mathrm{d}^3r'\;
\Phi_{\Lambda'}(\mathbf{r}')\,J_{\Lambda''}^{n}(\mathbf{r}',E_F),
\label{eq:density_matrix}
\end{align}
where the first term accumulates contributions from all energies up to $E_F$
through the scattering path operator $\tau_{\Lambda''\Lambda'''}^{nn}(E)$ and the
regular solutions $Z_{\Lambda}^{n}$, while the second term is an energy-independent
correction from the irregular solutions $J_{\Lambda''}^{n}$ ensuring proper
normalization within the atomic cell $\Omega_n$ \citep{Ebert2011}.

Both LDA$+U$ and DMFT introduce a self-energy $\Sigma_{\Lambda\Lambda'}$ acting
within the correlated subspace spanned by $\{\Phi_{\Lambda}\}$, and modify the
total energy and Green's function accordingly. They differ in whether this
self-energy is static or frequency-dependent.

\subsection{LDA\texorpdfstring{$+U$}{+U}}

Transforming $n_{\Lambda\Lambda'}^{n}$ from the relativistic $\Lambda=(\kappa,\mu)$
representation to the $L=(l,m,\sigma)$ representation, the LDA$+U$ total energy
functional reads \citep{Liechtenstein1995, Anisimov1997}:
\begin{equation}
E^{\mathrm{DFT}+U}(\rho, n)
=
E^{\mathrm{DFT}}(\rho)
+
E^{U}(n)
-
E^{\mathrm{dc}}(n),
\label{eq:ldau}
\end{equation}
where $E^{\mathrm{DFT}}(\rho)$ is the LSDA or GGA functional in terms of the spin
densities $\rho^{\uparrow}$ and $\rho^{\downarrow}$, $E^{U}(n)$ is the
electron-electron interaction energy of the localized electrons, and
$E^{\mathrm{dc}}(n)$ is the double-counting correction subtracting the Coulomb
interaction already contained in $E^{\mathrm{DFT}}(\rho)$. The double-counting
term is not uniquely defined; SPR-KKR supports both the fully localized limit and
the around-mean-field schemes \citep{Anisimov1997}. The corresponding static self-energy in the
$\Lambda$-subspace is
\begin{equation}
\Sigma_{\Lambda\Lambda'}^{U} =
\frac{\partial E^{U}(n)}{\partial n_{\Lambda'\Lambda}^{n}}
-
\frac{\partial E^{\mathrm{dc}}(n)}{\partial n_{\Lambda'\Lambda}^{n}},
\label{eq:sigma_ldau}
\end{equation}
which enters the Green's function as an energy-independent shift. This reduces
over-delocalization and improves the description of Mott insulators such as NiO
and CoO \citep{Imada1998}, but cannot capture dynamical effects such as Kondo
resonances or Hubbard satellites.

\subsection{DMFT}

For systems requiring dynamical correlations, heavy-fermion compounds, Kondo
systems, and materials near the Mott transition, the static self-energy
$\Sigma_{\Lambda\Lambda'}^{U}$ is replaced by a frequency-dependent counterpart
$\Sigma_{\Lambda\Lambda'}(E)$ obtained from a local quantum impurity problem
\citep{Georges1996}. The impurity is defined by the same correlated subspace
$\{\Phi_{\Lambda}\}$ and density matrix $n_{\Lambda\Lambda'}^{n}$ as in LDA$+U$,
ensuring a consistent projection throughout. The self-energy is projected back
onto the KKR basis as
\begin{equation}
\Sigma(E) = \sum_{\Lambda\Lambda'}
\Phi_{\Lambda}\,\Sigma_{\Lambda\Lambda'}(E)\,\Phi_{\Lambda'}^{\times},
\label{eq:dmft_sigma}
\end{equation}
where $\Phi_{\Lambda}^{\times}$ is the left-hand solution consistent with the KKR
biorthogonality convention of Eq.~(\ref{eq:green}), and modifies the lattice
Green's function through
\begin{equation}
\left[ E - H - \Sigma(E) \right] G(E) = \mathbf{1}.
\label{eq:dmft_G}
\end{equation}
The frequency dependence of $\Sigma_{\Lambda\Lambda'}(E)$ encodes Kondo resonances,
Hubbard satellites, and quasiparticle renormalization inaccessible to LDA$+U$
\citep{Georges1996, Kotliar2006}. Charge self-consistency between the KKR host
and the DMFT impurity solver \citep{Minar2005, Minar2011} ensures proper feedback
between local correlations and itinerant band states, essential for quantitative
spectroscopic predictions in strongly correlated systems \citep{Imada1998, Held2007}.
\section{Implementation Aspects Relevant for the Interface}
\label{sec:implementation}

SPR-KKR supports two potential approximations: the atomic sphere approximation (ASA)\citep{Andersen1975} , which assumes spherical potentials within touching spheres, and the full-potential mode \citep{Drittler1991}, which expands $V_i(\vec{r})=\sum_{\ell m}V_{i\ell m}(r)\,Y_{\ell m}(\hat{\vec{r}})$ at greater computational cost. ASA is efficient for close-packed structures, while the full-potential mode is necessary for open structures and directional bonding. For such open structures, empty spheres with zero nuclear charge can be inserted to improve basis completeness; ASE2SPRKKR provides automated empty-sphere generation.

Semi-infinite geometries for surface calculations \citep{MacLaren1990, Skriver1991} partition the system into a periodic bulk region and an interaction zone extending to vacuum. The screened (tight-binding) KKR variant \citep{Zeller1995, Szunyogh1994} truncates real-space sums beyond a cluster radius \texttt{CLURAD}, which is essential for large surface unit cells and reduces the $\mathcal{O}(N^3)$ matrix inversion cost significantly.

Task-specific calculations (SCF, DOS, BSF, ARPES, $J_{ij}$, XAS, and others) require distinct parameter sets organised hierarchically. ASE2SPRKKR exposes this structure through validated \texttt{InputParameters} objects while providing high-level interfaces for common workflows.

\bibliographystyle{elsarticle-num} 
\bibliography{references}